\documentclass[%
aip,jcp,amsmath,amssymb, reprint, twocolumn
]{revtex4-2}
\usepackage{amsmath}
\usepackage{mathtools}
\usepackage{float}
\usepackage{amsfonts}
\usepackage{amssymb}
\usepackage{dcolumn} %% tables cols aligned at decimal point
\usepackage{array}
\newcolumntype{P}[1]{>{\centering\arraybackslash}p{#1}}
\newcolumntype{M}[1]{>{\centering\arraybackslash}m{#1}}
\newcolumntype{C}[1]{>{\centering\arraybackslwash}p{#1}}
\usepackage{float}
\usepackage{psfrag}
\usepackage{tabularx}
\usepackage{stackengine}
\usepackage{amssymb}
\usepackage{mathtools}  
\usepackage{xfrac} 
\usepackage[T1]{fontenc}
\usepackage{graphicx}
\usepackage{diagbox}
\usepackage[caption=false]{subfig}
\usepackage{tikz}
\usetikzlibrary{positioning}
\usetikzlibrary{arrows}
\usetikzlibrary{trees}
\usepackage{amssymb}
\usetikzlibrary{decorations.pathmorphing}
\usetikzlibrary{decorations.markings}
\usetikzlibrary{automata,positioning}
\usepackage{braket}
\usepackage{rotating}
\usepackage{adjustbox}
\usepackage{ragged2e}
\usepackage{simplewick}
\usepackage{simpler-wick}

\usepackage{csquotes}
\usepackage{physics,amsmath}
\usepackage{xcolor}
\usepackage{fancyhdr}
\UseRawInputEncoding
\usepackage[normalem]{ulem}

\usepackage{scalerel}
\usepackage{hyperref}
\usepackage{appendix}

\begin{document}

\author{Chayan Patra}
\affiliation{ Department of Chemistry,  \\ Indian Institute of Technology Bombay, \\ Powai, Mumbai 400076, India}

% \author{Sonaldeep Halder}
% \affiliation{ Department of Chemistry,  \\ Indian Institute of Technology Bombay, \\ Powai, Mumbai 400076, India}

\author{Rahul Maitra}
\email{rmaitra@chem.iitb.ac.in}
\affiliation{ Department of Chemistry,  \\ Indian Institute of Technology Bombay, \\ Powai, Mumbai 400076, India}
\affiliation{Centre of Excellence in Quantum Information, Computing, Science \& Technology, \\ Indian Institute of Technology Bombay, \\ Powai, Mumbai 400076, India}
%\title{A Resource Efficient Projective Quantum Eigensolver via Adiabatically Decoupled Subsystem Evolution}
% \title{Toward a More Optimal Minima via Subspace Optimization and Non-iterative Corrections for Variational Quantum Algorithms}
%\title{Variational Quantum Eigensolvers via Subspace Optimization and Energy-landscape Plummeting for Improved Quantum Efficiency}
% \title{Variational Quantum Algorithms via Subspace Optimization and Non-iterative Corrections toward an Optimal Minima}
%\linenumbersCorrections
\title{Energy Landscape Plummeting in Variational Quantum Eigensolver: Subspace Optimization, Non-iterative Corrections and Generator-informed Initialization for Improved Quantum Efficiency}

\begin{abstract}

Variational Quantum Eigensolver (VQE) faces significant challenges due to hardware noise and the presence of barren plateaus and local traps in the optimization landscape.
To mitigate the detrimental effects of these issues, we introduce a general formalism that optimizes hardware resource utilization and accuracy by projecting VQE optimizations on to a reduced-dimensional subspace,
followed by a set of posteriori corrections.
Our method partitions the ansatz into a lower dimensional principal subspace and a higher-dimensional auxiliary subspace based on a conjecture of temporal hierarchy present
among the parameters during optimization.
The \textit{adiabatic approximation} exploits this hierarchy, restricting optimization to the lower dimensional principal subspace only.
This is followed by an efficient higher dimensional auxiliary space reconstruction without the need to perform variational optimization.
These reconstructed auxiliary parameters are subsequently included in the cost-function
via a set of auxiliary subspace corrections (ASC) leading to a \enquote{\textit{plummeting effect}} in the energy landscape toward a 
more optimal minima without utilizing any additional quantum hardware resources.
Numerical simulations show that, when integrated with any 
chemistry-inspired ansatz, our method can provide one to two orders 
of magnitude better estimation of the minima.
Additionally, based on the adiabatic approximation, we introduce a novel initialization strategy driven by unitary rotation generators for accelerated convergence of gradient-informed dynamic quantum algorithms.
Our method shows heuristic evidences of alleviating the effects of local traps, facilitating convergence toward a more optimal minimum.

\end{abstract}

\maketitle

\section{Introduction}
Variational quantum Eigensolvers (VQE)\cite{peruzzo2014variational, cerezo2021variational, bharti2022noisy} have emerged to be one of the most promising
hybrid algorithms that can potentially be used to achieve the quantum advantage in the Noisy Intermediate Scale Quantum (NISQ) era.
% In this framework, the optimization problem is formulated through a cost-function defined by the expectation values of observables evaluated on
% quantum states produced by a parameterized quantum circuit (PQC).
% A classical computer then iteratively adjusts the circuit's parameters to optimize these expectation values, effectively training the quantum circuit.
In this framework, the optimization problem is formulated through a cost-function defined by the expectation values of observables produced by a parameterized quantum circuit (PQC).
A classical computer then iteratively adjusts the circuit's parameters to optimize these expectation values, effectively training the quantum circuit.
% These parameters are optimized using classical processors.
Although VQEs are highly versatile and, in principle, capable of solving a wide range of classically
intractable problems, they face significant challenges when applied to complex tasks of practical relevance.
Among them the two major issues are the increasing circuit depth with problem size that often surpasses the current capabilities of NISQ devices and
the exponential decay of the cost function gradient with increasing qubit count, a phenomenon known as barren plateaus (BPs)\cite{mcclean2018barren, cerezo2021cost}.
In addition to BPs, another challenge in training arises from the prevalence of spurious local minima and traps in the cost function landscape
particularly when the PQC is too shallow as it does not contain enough parameters to explore all possible directions for optimization\cite{larocca2025barren,anschuetz2022quantum}.
% for a wide class of parameterized variational quantum models\cite{anschuetz2022quantum}.
Some recent theoretical developments\cite{larocca2023theory, kiani2020learning,wiersema2020exploring}
% by Larocca \textit{et al.}\cite{larocca2023theory}
suggest that
such local traps can be ameliorated by overparametrization of the associated PQCs, though the number of parameters
for this scales exponentially with the system size, making it an infeasible option.
To alleviate the effects of some of the above mentioned issues, adaptive-growth ansatz\cite{grimsley2019adaptive,Grimsley2023,tang2021qubit,yordanov2021qubit,zhu2022adaptive,cerezo2021variational,mondal2023development, halder2022dual,stair2021simulating,ryabinkin2018qubit,ryabinkin2020iterative} can be used.
It has been shown heuristically that some of these adaptive-growth ansatze can provide relatively shallower circuits and potentially bypass
local traps by scanning the parameter landscape locally via iteratively growing the ansatz in a gradient-informed manner\cite{Grimsley2023}.
Despite their effectiveness and widespread adoption,
these protocols still demand extensive measurement overhead
and large number of parameters (and quantum gates) to navigate through
many local minimas in larger systems to achieve desired accuracy,
posing significant challenges for scalability.
In this regard, striking the perfect balance between computational 
resource-efficiency (such as circuit depth and measurement) and accuracy, along
with better parameter initialization still remain as open questions 
and are active areas of research in recent times\cite{yordanov2020efficient,magoulas2023cnot,mondal2024projective,patra2024toward,halder2023corrections,halder2024noise,halder2024machine,kowalski2021dimensionality,kowalski2023quantum,robledo2024chemistry,mcclean2018barren,bittel2021training,holmes2022connecting,grant2019initialization,wang2024trainability}.

In this work, we take a step forward toward the optimization of this trade-off with a general formalism that maintains the resource-efficiency by
including minimal number of parameters into the optimizable PQC (ansatz) while the desired accuracy
of the cost-function is ensured by a set of one-step non-iterative corrections.  
These non-iterative corrections do not require any additional quantum hardware resources or classical optimization routines.
To achieve this, our formalism conjectures a temporal hierarchy among the variational parameters regarding the \enquote{timescale} of convergence to
decouple the ansatz into a dominant lower-dimensional \textit{principal} subspace and a submissive higher-dimensional \textit{auxiliary} subspace.
% Based on some recent works from our group\cite{patra2024toward, patra2024projective, halder2023machine, agarawal2020stability, agarawal2021accelerating, agarawal2021approximate, patra2023synergistic}, which reveal coupled-cluster\cite{bartlett2007coupled, crawford2000introduction} (CC)-like parameter optimizations
% evolve on markedly different \enquote{timescales} (or, the number of iterations for convergence),
% we conjecture that
% there exists a temporal hierarchy among the variational parameters.
Consequently, we invoke the \textit{adiabatic approximation}\cite{patra2024toward, patra2024projective, halder2023machine, agarawal2020stability, agarawal2021accelerating, agarawal2021approximate, patra2023synergistic} to decouple the parameter set by neglecting the successive variations of the auxiliary parameters
in the characteristic timescale of principal ones and the optimization is cast on to a reduced dimensional subspace.
Such principal parameters selection may be achieved by threshold-based perturbative
estimates (like second-order M{\o}ller-Plesset or MP2 estimate for molecular systems) or gradient measurements (like using Adaptive, Problem-Tailored VQE or ADAPT-VQE\cite{grimsley2019adaptive, Grimsley2023}). In the former case,
the selection of operators is independent of hardware noise and does not get corrupted by the noise profile.
% The principal subspace can be selected by
% any variable-structure ansatz construction protocol\cite{grimsley2019adaptive,mondal2023development,halder2022dual}
% based on some threshold criterion
% of the associated gradient measurements. In currently available noisy hardware, often measurement-based operator selection is unreliable as
% noise can affect the selection of relevant operators\cite{robledo2024chemistry,kanno2023quantum,halder2024machine}. To maintain
% the structural sequence of many-body operators even under noise,
% we can opt for a measurement-free selection protocol\cite{halder2024noise,patra2024projective} that is based on classically
% efficient-to-compute perturbative
% estimates of the corresponding parameters.
While more relaxed threshold conditions are often better for compact gate-efficient circuits,
it does not always minimize the cost function to the desired accuracy and can get stuck in far-from-optimal local minimas.
In such a scenario more gradient-informed parameters must be included (from the auxiliary subset according to our language)
in the PQC for the \enquote{burrowing} effect\cite{Grimsley2023} of the parameter landscape toward a better minima.
However, the direct inclusion of these parameters into the PQC inevitably increases circuit depth, exacerbating resource utilization.
To circumvent this, we propose a framework that can efficiently reconstruct the auxiliary parameters from the optimized
principal parameters without the need to perform variational optimization and incorporates the effect of these parameters into the cost-function through one-step posteriori
corrections.
Starting from the \textit{parameter-shift rule}\cite{mitarai2018quantum, schuld2019evaluating} for gradient evaluations, we establish a mathematical relationship that can reconstruct the auxiliary
parameters from the optimized principal parameters. We call this the \textit{principal-to-auxiliary mapping}
technique which is a direct consequence of the \textit{adiabatic approximation}.
These predicted auxiliary parameters can be incorporated approximately into the cost-function via a set of non-iterative corrections, termed as \textit{auxiliary subspace corrections}
(ASC) which induce a \enquote{plummeting effect} in the energy landscape toward a more optimal minima. The computations regarding the principal-to-auxiliary mapping and the subsequent ASC
can be executed in parallel quantum processors
without requiring any additional hardware resources.
The additional overhead due to ASC includes only a nominal number of extra quantum measurements which is bounded by the total number of operators in the pool.
Since no extra parameters are required into the PQC to perform these extra calculations,
the maximum circuit-depth of the entire protocol is governed by the low-dimensional principal subspace.
Our method is called \textit{adiabatically decoupled \enquote{X} with auxiliary subspace corrections} or AD(X)-ASC, where
\enquote{X} is the user-defined protocol used to select the principal subspace.

The paper is structured as follows: Section \ref{theory} establishes the theoretical framework, including adiabatic approximation, principal-to-auxiliary mapping and ASC. Section \ref{principal space selection} covers
possible choices regarding efficient principal subspace selection, particularly for electronic structure problems.
In section \ref{ASC measurement overhead} we discuss the computational overhead and bounds for ASC.
To numerically validate our approach, we apply it to electronic structure problems
in Section \ref{result and discussion}
using different tools such as ADAPT-VQE\cite{grimsley2019adaptive, Grimsley2023} or MP2 screening to select the principal subspace.
For noiseless cases, our method yields up to two orders-of-magnitude accuracy improvement compared to ADAPT-VQE
across the PES without additional quantum resources. Noisy simulations further confirm significant energy estimation improvements due to ASC.
In section \ref{initialization technique} we introduce a generator-driven initialization based on the adiabatic approximation
which reduces quantum measurement costs by up to 48\%.

\section{Theory} \label{theory}
% \section{VQE as a multi-variable \enquote{dynamical} system} \label{general principle section}
\subsection{Parameter Space Decoupling in VQE based on Temporal Hierarchy and Dimensionality Reduction
via Adiabatic Approximation} \label{general principle section}
Variational quantum eigensolvers (VQE) \cite{peruzzo2014variational} resort to a quantum-classical hybrid architecture that
encode the problem of interest in
a cost function of the form
\begin{equation}
    f(\theta) = \bra{\Phi_0}\hat{U}(\boldsymbol{\theta})^\dagger \hat{B} \hat{U}(\boldsymbol{\theta})\ket{\Phi_0}
\end{equation}
where, $\boldsymbol{\theta}$ is a set of $N_{par}$ real gate parameters
encoded in the unitary operator $\hat{U}(\boldsymbol{\theta})$, $\ket{\Phi_0}$ is a reference state and $\hat{B}$ is a bounded, self-adjoint operator.
Due to the parameterized nature of $\hat{U}(\boldsymbol{\theta})$, the terms unitary operator and parameter are used interchangeably in the manuscript.
VQE optimizes the cost function to find the minima of the landscape that provides the solution of the associated
problem.
The unitary operator $\hat{U}(\boldsymbol{\theta})$ typically contains a sequence of quantum gate
operations having the disentangled form
\begin{equation}
    \hat{U}(\boldsymbol{\theta}) = \prod_{\mu} \hat{U}_{\mu} (\theta_\mu).
\end{equation}
with
% \begin{equation} \label{U_mu general}
%     \hat{U}_\mu (\theta_\mu) = \prod_l e^{-i \theta_l \hat{G}_l} \hat{W}_l.
% \end{equation}
\begin{equation} \label{U_mu general}
    \hat{U}_\mu (\theta_\mu) =  e^{-i \theta_\mu \hat{G}_\mu} \hat{W}_\mu.
\end{equation}
Here, $\hat{W}_\mu$ is an unparameterized unitary and $\hat{G}_\mu$ is a Hermitian operator.
In the language of Lie algebra, $\hat{G}_\mu$ is called the generator of the unitary $\hat{U}_\mu(\theta_\mu)$.
Since the theoretical framework we are going to develop involves only the parameterized part of the unitary,
from here onwards for mathematical simplicity we set the $\hat{W}_\mu$ operators to be identity operators
without sacrificing any generality or conceptual jargon. However, it can always be included into the ansatz depending on the problem at hand
and the subsequent analysis would be the same with some extra terms involving $\hat{W}_\mu$s.
In general, VQE, which is grounded in the variational principle, collects the expectation values of the cost function from
a quantum device and outsource it to a classical processing unit for optimization.
Usually classical optimizers perform numerical differentiation to obtain approximate gradients
or in some cases gradient-free optimizers are also preferred specially under noisy scenarios.
However, the gradients of the cost function $f(\boldsymbol{\theta})$ with respect to $i$-th parameter $\theta_i$ can also be calculated in a quantum device
by the \textit{parameter-shift rule} \cite{mitarai2018quantum, wierichs2022general}
\begin{equation} \label{parameter shift rule generalized}
    \frac{\partial f(\boldsymbol{\theta})}{\partial \theta_i} = \frac{f(\boldsymbol{\theta}+s\hat{e}_i) - f(\boldsymbol{\theta}-s\hat{e}_i)}{2 \sin{(\Omega s)}} \Omega
\end{equation}
where, $\hat{e}_i$ is a unit vector with all the zero elements except $i$-th element (which is 1), $s$ is the shift in the gate parameter and
$\Omega$ is the positive unique differences of the eigenvalues of the corresponding generator $\hat{G}_l$.
If $\hat{G}_l$ is a linear combination of Pauli operators, it has two unique eigenvalues and in such cases $\Omega=1$ and
$s$ is often taken to be $\frac{\pi}{2}$ leading to a widely used expression of the parameter shift rule\cite{schuld2019evaluating}
\begin{equation} \label{parameter shift rule}
    \frac{\partial f(\boldsymbol{\theta})}{\partial \theta_i} = \frac{1}{2} \Big[f(\boldsymbol{\theta}+\frac{\pi}{2}\hat{e}_i) - f(\boldsymbol{\theta}-\frac{\pi}{2}\hat{e}_i)\Big]
\end{equation}
Unlike numerical differentiation, the parameter-shift rule provides an exact analytical expression of the gradients which has
a stochastic implementation\cite{schuld2019evaluating}.
Considering the analytical gradient expression of Eq. \eqref{parameter shift rule}, one can write the parameter update equation as
\begin{equation}\label{theta_m+1}
    \theta_i^{(m)} = \theta_i^{(m-1)} + \alpha \frac{\partial f(\boldsymbol{\theta})}{\partial \theta_i} \hspace{5mm} \forall i
\end{equation}
where, $\alpha$ is some constant usually called the learning-rate and $m$ is the iteration step.
Re-ordering Eq.\eqref{theta_m+1} leads to $\Delta \theta_i^{(m)} = \theta_i^{(m)} - \theta_i^{(m-1)}$ at $m$-th step as
\begin{equation}\label{delta theta}
    \Delta \theta_i^{(m)} = \alpha \frac{1}{2}\Big[f(\boldsymbol{\theta}+\frac{\pi}{2}\hat{e}_i) - f(\boldsymbol{\theta}-\frac{\pi}{2}\hat{e}_i)\Big]
\end{equation}
where we have used Eq.\eqref{parameter shift rule} to substitute $\frac{\partial f}{\partial \theta_i}$.
Note that due to the presence of the parameter vector $\boldsymbol{\theta}$ in the update equation
for $\theta_i$, the optimization involves a set of coupled difference equations which are highly nonlinear in nature.

Generally in such multivariate optimization problems, different parameters reach the fixed point in different timescales
if we map iteration steps into a discretely spaced time domain\cite{agarawal2021approximate, patra2023synergistic, van1985elimination,halder2023machine,patra2024projective, patra2024toward}. To proceed further, for all practical scenarios we assume that we can classify the entire parameter
set into two broad categories based on their optimization pattern: (1) Auxiliary Parameters ($\boldsymbol{\theta_A}$): these $N_A$ number of parameters are damped modes and converge faster, or in other words
they take fewer number of iterations to reach the fixed point, (2) Principal Parameters ($\boldsymbol{\theta_P}$): this subset contains $N_P$ number of parameters and they usually take relatively more number of iterations to converge. In general, the principal parameter
subset is a much lower dimensional subspace than
its auxiliary counterpart such that $N_P << N_A$ and as the name suggests, this subset usually governs the entire optimization trajectory.
Thus, our target is to cast the entire parameter optimization task into this 
lower dimensional principal subspace.
Such parameter space decoupling directly implies the corresponding unitary space to be decoupled into an auxiliary and principal unitary
\begin{equation} \label{U = U_P U_A}
    \hat{U} = \prod_{A_i} \hat{U}_{A_i} \prod_{P_I} \hat{U}_{P_I}
\end{equation}
where, $\hat{U}_{P_I}$ ($\hat{U}_{A_i}$) are principal (auxiliary) unitaries which have the form of Eq. \eqref{U_mu general}
with $\mu$ being replaced by principal (auxiliary) indices $P_I$ ($A_i$) and are to be treated on different footing (discussed later).
Under such a decoupling, the associated cost function becomes
\begin{equation} \label{cost function under aidbatic approx}
    f = \bra{\Phi_0} (\prod_{P_I} \hat{U}_{P_I})^{\dagger} (\prod_{A_i} \hat{U}_{A_i})^{\dagger} \hat{B} \prod_{A_i} \hat{U}_{A_i} \prod_{P_I} \hat{U}_{P_I} \ket{\Phi_0}
\end{equation}
One must note that since such unitaries do not in general commute, in Eq. \eqref{U = U_P U_A} the ordering depends heuristically on the specific problem
and the definition of such a decoupling of unitary is flexible according to the need of the problem at hand.
% For example selected projective quantum eigensolver\cite{stair2021simulating} (SPQE) and ADAPT-VQE\cite{grimsley2019adaptive}
% algorithms follow different operator orderings when
% it comes to optimizing the structure of an ansatz.
% Such an assumption lay the ground for our further analysis
% and in later sections we will substantiate it with numerical examples.

The assumptions discussed so far indicate that the cost function reaches a fixed point in the characteristic convergence timescale of the principal parameters. Throughout the optimization process, since the auxiliary parameters
converge relatively much faster compared to the principal ones, the variation of
auxiliary parameters between two consecutive iterations is negligible. Thus the iterative variations for auxiliary parameters can safely be ignored in the characteristic timescale of principal parameter convergence
\begin{equation} \label{adiabatic approx}
    \Delta \theta_{A_i} = 0 \hspace{5mm} \forall A_i .
\end{equation}
We term this the \textit{adiabatic approximation}. 
With this approximation, the optimization task is restricted only to the lower dimensional principal subspace
\begin{equation}
    \theta_{P_I}^{(m+1)} = \theta_{P_I}^{(m)} + \alpha \frac{\partial f(\boldsymbol{\theta_P})}{\partial \theta_{P_I}} \hspace{2mm} \forall P_I 
\end{equation}
which utilizes much reduced quantum resources compared to the optimization of the entire parameter space. Another advantage of effectively encoding the
optimization dynamics in a smaller subspace is that it may be less prone to be affected by barren plateaus\cite{larocca2025barren}.
However, often a shallow PQC (in our case parametrized by $\boldsymbol{\theta_P}$) suffers from spurious local minimas as it
explores restricted low dimensional manifolds of the entire cost-function hypersurface.
In general, increasing the number of parameters in the PQC enhances variational flexibility, allowing the optimizer to explore more available directions toward optimal minimum.
This of course comes at the cost of increasing the circuit depth and hence one must strike the right balance to handle this delicate trade-off. 
% Generally, by increasing the number of parameters in the PQC one can
% introduce more variational flexibility to explore more available direction towards an optimal minima during the optimization.
% during this principal optimization, the auxiliary parameters are entirely ignored which can have important correlation
% effects.
Toward this in the next subsection we introduce a set of auxiliary subspace corrections that incorporate the effects of all the parameters of the pool
into the cost function without explicitly including them into the PQC.
% In the next subsection, we will discuss how we can incorporate the effects of the auxiliary parameters into the cost function efficiently.

\subsection{Principal-to-Auxiliary Functional Mapping and Higher-Order Auxiliary Subspace Correction to the Cost Function}

Due to the presence of nonlinear coupling among the parameters, even if the explicit variation of the
auxiliary parameters are frozen via adiabatic approximation, certain implicit effects of the principal
subset should be reflected upon the auxiliary set.
Quantitatively, one can consider Eq. \eqref{delta theta} for auxiliary parameters and invoke the adiabatic approximation
(Eq. \eqref{adiabatic approx}), leading to
\begin{equation} \label{delta theta_A}
    \Delta \theta_{A_i} = \frac{1}{2}\Big[f(\boldsymbol{\theta}+\frac{\pi}{2}\hat{e}_{A_i}) - f(\boldsymbol{\theta}-\frac{\pi}{2}\hat{e}_{A_i})\Big] = 0
\end{equation}
where the cost-function $f$ has the form of Eq. \eqref{cost function under aidbatic approx}.
The expectation value terms in Eq. \eqref{delta theta_A} can be analytically evaluated as
\begin{equation} \label{f_theta expansion}
\begin{split}
    &f(\boldsymbol{\theta} \pm \frac{\pi}{2}\hat{e}_{A_i}) = \bra{\Phi_0}\Big(\prod_{P_I} \hat{U}_{P_I}\Big)^{\dagger} . . . \hat{U}_{A_i}^{\dagger}(\theta_{A_i} \pm \frac{\pi}{2}). . . \\
    & \hat{U}_{A_2}^{\dagger} \hat{U}_{A_1}^{\dagger} \cdot \hat{B} \cdot \hat{U}_{A_1} \hat{U}_{A_2} ... \hat{U}_{A_i}(\theta_{A_i} \pm \frac{\pi}{2}) ... \prod_{P_I} \hat{U}_{P_I} \ket{\Phi_0}
\end{split}
\end{equation}
Here, the auxiliary part of the unitary is expanded to explicitly reflect the parameter shift in the
$A_i$-th component. By employing the Baker-Campbell-Hausdorff (BCH) expansion for auxiliary unitaries, terms up to second order in $\theta_{A_i}$
are retained, resulting in the approximate function (see Appendix \ref{cost function approx appendix} for details)
\begin{equation}\label{f_theta_plus_minus_pi_by_2_upto_second_order}
\begin{split}
    & f(\boldsymbol{\theta}\pm\frac{\pi}{2}\hat{e}_{A_i}) \approx \bra{\Phi_P} \hat{B} \ket{\Phi_P} + (\theta_{A_i}\pm\frac{\pi}{2}) \\
    & \bra{\Phi_P}[\hat{B}, \hat{\mathcal{G}}_{A_i}] \ket{\Phi_P} + \sum_{A_{j \neq i}} \theta_{A_j} \\
    &  \bra{\Phi_P}[\hat{B}, \hat{\mathcal{G}}_{A_j}]\ket{\Phi_P} + \frac{1}{2} (\theta_{A_i}\pm\frac{\pi}{2})^2 \\
    & \bra{\Phi_P}\Big[ [\hat{B}, \hat{\mathcal{G}}_{A_i}], \hat{\mathcal{G}}_{A_i} \Big] \ket{\Phi_P} 
    % + \frac{1}{2} \sum_{A_j} \sum_{A_k\ne A_i} \theta_{A_j} \theta_{A_k} \\
    % &\bra{\Phi_P}\Big[ [\hat{B}, \hat{\mathcal{G}}_{A_j}], \hat{\mathcal{G}}_{A_k} \Big] \ket{\Phi_P}
    % + \sum_{A_j, A_k} \theta_{A_j} \theta_{A_k} \bra{\Phi_P}\Big[ [\hat{B}, \hat{\mathcal{G}}_{A_j}], \hat{\mathcal{G}}_{A_k} \Big] \ket{\Phi_P}
\end{split}    
\end{equation}
Here, $\hat{\mathcal{G}}_{A_j}= - i \hat{G}_{A_j}$ is an anti-Hermitian operator and
\begin{equation} \label{phi_P}
    \ket{\Phi_P} = \prod_P \hat{U}_P (\theta_P) \ket{\Phi_0}.
\end{equation}
With the expression in Eq. \eqref{f_theta_plus_minus_pi_by_2_upto_second_order} of the cost function, 
Eq. \eqref{delta theta_A} may be simplified to (see Appendix \ref{principal to auxiliary mapping from params shift rule appendix} for the derivation and the 
approximations)
\begin{equation}\label{adiabatic_approx_theta_A}
    0 =  \bra{\Phi_P} [\hat{B}, \hat{\mathcal{G}}_{A_i}] \ket{\Phi_P} + \theta_{A_i} \bra{\Phi_P} \Big[[\hat{B}, \hat{\mathcal{G}}_{A_i}], \hat{\mathcal{G}}_{A_i} \Big] \ket{\Phi_P}
\end{equation}
This leads to the \textit{principal-to-auxiliary functional mapping} where 
$\theta_{A_i}$ is reconstructed from the composite set of optimized principal parameters:
\begin{equation} \label{theta_A as function of theta_P}
    \theta_{A_i} (\boldsymbol{\theta_P}) = - \frac{\bra{\Phi_P}[\hat{B}, \hat{\mathcal{G}}_{A_i}]\ket{\Phi_P}}{\bra{\Phi_P}\Big[ [\hat{B}, \hat{\mathcal{G}}_{A_i}], \hat{\mathcal{G}}_{A_i} \Big]\ket{\Phi_P}}
\end{equation}
% This expression is particularly significant, as it establishes a
% functional relationship to reconstruct each of the auxiliary parameters
% from the set of principal parameters.
With this recipe to reconstruct the auxiliary parameters from the 
principal ones, the optimization (using Eq. \eqref{theta_m+1}) can be
restricted to the lower dimensional principal parameter subspace only.
The auxiliary parameters can subsequently be reconstructed by substituting 
only the optimized principal parameters into 
Eq. \eqref{theta_A as function of theta_P} in a single step and thus avoiding the variational optimization
for auxiliary subspace.

With this formulation, the corresponding expression for the optimized cost function is given by
\begin{equation} \label{final cost function 1}
    \begin{split}
        f \equiv f(\theta_P) &= \bra{\Phi_0} (\prod_{P_I} \hat{U}_{P_I}(\theta_{P_I}^{(opt)}))^{\dagger} (\prod_{A_i} \hat{U}_{A_i}(\theta_{A_i}(\boldsymbol{\theta_P}))^{\dagger} \hat{B}\\
        & \prod_{A_i} \hat{U}_{A_i}(\theta_{A_i}(\boldsymbol{\theta_P})) \prod_{P_I} \hat{U}_{P_I}(\theta_{P_I}^{(opt)}) \ket{\Phi_0}
    \end{split}
\end{equation}
where, the operators $\hat{U}_P$ and $\hat{U}_A$ contain the optimized principal parameters $\theta_P^{(opt)}$
and mapped auxiliary parameters $\theta_A(\boldsymbol{\theta_P})$ respectively.
However, direct implementation of the cost function of the form 
in Eq. \eqref{final cost function 1} is not gate-efficient as it requires deep
circuit involving both optimized principal and mapped auxiliary
parameters in the PQC. To circumvent the construction of such 
deep circuits,
we can treat principal unitaries exactly and break the auxiliary unitary part $(\prod_A \hat{U}_A)^{\dagger} \hat{B} \prod_A \hat{U}_A$
according to BCH expansion while keeping only upto second order terms
in $\theta_{A_i}$
% (see Appendix \ref{principal to auxiliary mapping from params shift rule appendix})
\begin{equation} \label{cost function with ASC}
\begin{split}
    f \approx & \bra{\Phi_P} \hat{B} \ket{\Phi_P} + \sum_{A_i} \theta_{A_i}(\boldsymbol{\theta_P}) \bra{\Phi_P}[\hat{B}, \hat{\mathcal{G}}_{A_i}]\ket{\Phi_P} \\
    & + \frac{1}{2} \sum_{A_i} (\theta_{A_i}(\boldsymbol{\theta_P}))^2 \bra{\Phi_P}\Big[ [\hat{B}, \hat{\mathcal{G}}_{A_i}], \hat{\mathcal{G}}_{A_i} \Big]\ket{\Phi_P} 
\end{split}
\end{equation}
It is to be noted that in Eq. \eqref{f_theta_plus_minus_pi_by_2_upto_second_order} and \eqref{cost function with ASC}, we retain only the double commutator
terms of the form $\Big[ [\hat{B}, \hat{\mathcal{G}}_{A_i}], \hat{\mathcal{G}}_{A_i} \Big]$
while disregarding all cross terms. This choice is motivated by the fact that incorporating cross terms
would significantly increase the number of additional measurements.
Moreover, during our numerical analysis, we observed
the contributions arising from the cross terms are negligible compared to other leading order terms in the series, justifying their omission.
The detailed mathematical steps regarding the approximations are discussed in details in Appendix \ref{cost function approx appendix} and \ref{principal to auxiliary mapping from params shift rule appendix}.
The first term of Eq. \eqref{cost function with ASC} represents the iteratively optimized component of the
cost function which involves only the principal subspace whereas the subsequent terms are higher order
corrections stemming from the BCH expansion involving the auxiliary parameters.
We call these higher order corrections to the cost function as the \textit{auxiliary subspace correction} (ASC)
terms. As we will discuss in section \ref{adapt for principal selection} that the principal subspace can be
selected via any efficient operator selection protocol (say it is called \enquote{X}) followed by the subsequent
ASC. Thus the entire framework developed so far in general is to be called \textit{adiabatically decoupled \enquote{X} method with
auxiliary subspace corrections} or AD(X)-ASC.
The advantage of our method is that the circuit complexity for the entire process that involves the parameter optimization, principal-to-auxiliary 
subspace mapping and ASC is governed by \enquote{X} only and thus the posteriori 
correction due to ASC comes with almost no additional cost.
Furthermore, AD(X)-ASC is a plug-and-play algorithmic model where the 
choice of the method $X$ is user defined and can be chosen from any dynamic or static ansatz engineering model.
In this context, it is important to mention that there have been some recent works on non-iterative corrections\cite{claudino2021improving,magoulas2023unitary,windom2024new, haidar2024non}
for energy improvements based on perturbative inclusions or cluster moment expansions. Subalgebra inspired decoupled structure of the ansatz have also been studied extensively
for reduced resource requirements in the Double Unitary CC (DUCC) framework in both classical\cite{kowalski2018properties, kowalski2018regularized} and quantum
computing\cite{kowalski2021dimensionality, kowalski2023quantum} domain.
On the contrary, AD(X)-ASC
relies upon the adiabatic approximation for ansatz decoupling while the non-iterative ASC stem from the parameter-shift rule with judiciously truncated BCH expansions.
This provides a general mathematical framework which,
we believe, would enable wide applications of the algorithm.

To numerically substantiate the advantages of theoretical framework developed so far, we will consider the
electronic structure problems of molecules under the Born-Oppenheimer approximation.
% case of molecular electronic
% structure calculations.
Since molecular electronic structure theory in the quantum computing framework has a well-defined problem-inspired disentangled
unitary coupled cluster (dUCC) ansatz\cite{dUCC_evangelista}, we need to make the necessary replacements in our general framework discussed
in section \ref{general principle section}. Regarding this, we will set the unitary in Eq. \eqref{U_mu general} to be
\begin{equation}
    \hat{U}_\mu = e^{\theta_\mu \hat{\kappa}_\mu}
\end{equation}
with the operator $\hat{B}$ replaced by molecular Hamiltonian $\hat{H}$,
rotation generators $\hat{\mathcal{G}}_{\mu}$ by the anti-hermitian operator $\hat{\kappa}_\mu$ where $\hat{\kappa}_\mu = \hat{\tau}_\mu - \hat{\tau}_\mu^{\dagger}$
with $\hat{\tau}_\mu$ being the coupled-cluster excitation operator defined by $\hat{\tau}_\mu = \hat{a}_a^{\dagger} \hat{a}_b^{\dagger} ... \hat{a}_j \hat{a}_i$.
Here $\hat{a}_a^\dagger$ ($\hat{a}_i$) is a particle creation (annihilation) operator defined on the Hartree-Fock reference state $\ket{\Phi_0}$ and $\mu$ is a composite
hole-particle index.
In section \ref{principal space selection} we will discuss
some efficient choices of X to select the principal parameter subspace.

% In the next sections, we will
% analytically and numerically analyze two cases where $X$ is chosen to be (a) ADAPT-VQE as the principal subspace selection 
% technique and (b) standard UCCSD ansatz where the parameters are filtered in based on MP2 amplitudes. 

\section{Construction of the Principal Subspace Unitary} \label{principal space selection}

% The choice of the principal parameter subset is extremely important as ultimately it governs the entire optimization dynamics. As we discussed in section
% \ref{general principle section} this subset usually consist of the slower converging parameters that have larger variations in each successive iterations.
The efficient selection of the principal parameter subspace and the ordering of the 
corresponding operators are of paramount importance and is orchestrated by method \enquote{X}.
The choice of \enquote{X} thus fundamentally dictates the optimization and 
overall resource requirements. As elaborated in Section \ref{general principle section}, this subset typically comprises parameters that exhibit slower convergence and undergo more substantial variations across successive iterations. While the principal subspace can be formed via any user-defined protocol,
in Section \ref{adapt for principal selection} and \ref{mp2 for principal selection} we will discuss two possible efficient choices for the same.

% \textcolor{red}{WRITE THAT THE ORDERING OF UP depends on X.}

% In the next few sections we will discuss how the way different popular variable-structure ansatze basically selects \enquote{important} operators are implicitly
% based upon such a timescale decoupling of the parameter space and hence these techniques are naturally the preferable tool
% for the principal parameter selection.

%The notion of Adiabatic Decoupling in ADAPT-VQE and Principal Subset Selection
\subsection{ADAPT-VQE Based Selection of Operators for Principal Subspace} \label{adapt for principal selection}
% Adaptive Derivative-Assembled Pseudo Trotterized VQE or 
ADAPT-VQE \cite{grimsley2019adaptive, Grimsley2023}
has been one of the most popular adaptive-growth ansatz construction protocols in VQE framework. 
% Fixed structured ansatz often gives control over the overall circuit complexity while it can miss out on circuit structure optimization.
% by adding only problem-specific
% important elements or discarding unnecessary ones.
% Variable structure ansatz like ADAPT-VQE\cite{grimsley2019adaptive}
% It can optimize the circuit structure by adaptively
% adding only problem-specific
% important elements or discarding unnecessary ones.
It consists of two alternate cycles: macro and micro-iteration cycles.
In the $k$-th macro-iteration step, it examines
the numerical values of all the gradients of the cost-function with respect to individual parameters of the form
% To minimize the proliferation of the circuit-depth caused by the presence less-important operators in the
% ansatz, it
% sets an \enquote{importance} criteria for selective inclusion of the operators in the ansatz from the entire pool of operators.
% Quantitatively, the approximate gradients of the cost function at iterative step $k$ is given by
\begin{equation}\label{adapt gradient equation}
   \frac{\partial f(\boldsymbol{\theta})}{\partial \theta_i} \approx \bra{\Phi_0} \hat{U}^{\dagger(k)} [\hat{H}, \hat{\mathcal{G}}_{i}] \hat{U}^{(k)} \ket{\Phi_0}; \forall i.
\end{equation}
It then appends the operator with largest gradient to $\hat{U}(k)$ in an one-operator-at-a-time manner
with the corresponding parameter initialized to zero.
The already added parameters from the previous ($k-1$) macro-iterations are
initialized using their optimal values obtained from the ($k-1$)-th step.
This is known as the \textit{recycled initialization} technique\cite{Grimsley2023}.
% ADAPT-VQE consists of two alternate cycles: macro and micro-iteration cycles. 
This is followed by the micro-iterative cycle that optimizes all the corresponding $k$ number of parameters via VQE (according to Eq. \eqref{theta_m+1})
to construct the trial ansatz $\hat{U}^{(k+1)} \ket{\Phi_0}$ for the $(k+1)$-th iteration of
the operator selection process.

The entire process is governed by a threshold $\epsilon$ and the iteration is terminated if at a particular macro-iteration $k$
all the gradients are smaller than $\epsilon$:
% the following condition is satisfied:
\begin{equation} \label{adapt convergence criteria}
    \bra{\Phi_0} \hat{U}^{\dagger(k)} [\hat{B}, \hat{\mathcal{G}}_{i}] \hat{U}^{(k)} \ket{\Phi_0} < \epsilon \mbox{      } ;\forall i.
\end{equation}
% In such a way ADAPT-VQE forms an ansatz by gradient-informed, one-operator-at-a-time manner which is compact
% and converges quickly to very accurate solutions.
The selection of the parameters in ADAPT-VQE with largest gradient shows it only
picks up those parameters that have the largest variation and hence takes comparatively more iterations to converge
than others since (see Eq. \eqref{theta_m+1}, \eqref{delta theta})
\begin{equation}
    \Delta \theta_i \propto  \frac{\partial f}{\partial \theta_i}
\end{equation}
% This is indeed the case for PQE based variable structure ansatz as well as discussed in some of our previous
% works\cite{patra2024projective, patra2024toward}.
where, $\Delta{\theta_i}$ is the successive variation of the parameter $\theta_i$.

This particular largest gradient-based selection criteria indicates that
it essentially follows the adiabatic decoupling principle at each step of the 
operator selection and thus the selected parameters are considered here to constitute 
the principal unitary. 
This makes ADAPT-VQE one of the most efficient
tools for the principal subspace selection and optimization.
Since the macro-iteration termination condition (Eq. \eqref{adapt convergence criteria}) has to be satisfied by all operators in the pool,
% The leftover \enquote{not-so-important}
the entire operator pool can be placed into the auxiliary subset post this termination
as $\Delta \theta_i$ are extremely small $\forall i$ at termination.
% This is due to the fact that
% ADAPT-VQE continuously scans and modifies the optimization landscape in each iteration
% % and allows repetition of operators in the ansatz
% and at $m$-th iteration, the gradient of the cost-function with respect to all the operators in the pool
% is much less compared to gradients at iterations $<m$.
The auxiliary parameters can be obtained post ADAPT-VQE optimization via Eq. \eqref{theta_A as function of theta_P}
and it will ultimately provide the one-step ASC to the cost-function.
However, since ADAPT-VQE relies upon measurement-based selection, which is accurate under noiseless circumstances,
the state preparation and measurement (SPAM) errors in real hardware can corrupt this operator selection\cite{robledo2024chemistry}.
In such a scenario, we can opt for the selection protocol discussed in the next section \ref{mp2 for principal selection}.

\subsection{Threshold Based Measurement-free Selection of Operators for Principal Subspace} \label{mp2 for principal selection}

An alternative approach regarding the selection of the principal subset
involves decoupling the parameter space based on initial guesses
obtained from computationally inexpensive classical methods.
Particularly for molecular Hamiltonians, some of the present authors have shown in Ref.\cite{agarawal2021approximate}
that the parameters with larger absolute magnitude from the second-order M{\o}ller-Plesset perturbative (MP2) estimates
take much more iterations to converge than the ones with smaller magnitude for CC optimization.
In this case MP2 parameters can be easily obtained from a classical computer
owing to its relatively reduced scaling $(\sim \mathcal{O}(N^5))$ with the system size $N$.
We define a threshold $\bar{\epsilon}$ and include
only those MP2 parameters from the entire set $\{\theta^{(MP2)}\}$ into the principal parameter subset for which
$\mid \theta^{(MP2)}_P \mid > \bar{\epsilon}$, where the subscript $P$ indicates principal subspace.
% The remaining parameters $\{\theta^{(MP2)}_A\}$ form the auxiliary subspace. 
% can be achieved via
% the second-order M{\o}ller-Plesset perturbative (MP2) estimates of the parameters.
Along with screened MP2 parameters, in the principal subspace we also include parameters corresponding to single excitations within the same spin sector.
The corresponding ansatz is defined as
% largest MP2-screened double excitations
% act on the HF reference first followed by smaller 
\begin{equation} \label{mp2s vqe ansatz}
    \hat{U}_P\ket{\Phi_0} = \prod_{\sigma=N_S}^{1} \hat{U}_P^{(S_\sigma)}(\theta^{S_\sigma}_P) \prod_{\tau=N_D}^1 \hat{U}_P^{(D_\tau)}(\theta^{D_\tau}_P) \ket{\Phi_0}
\end{equation}
where, $\hat{U}_P^{S/D}(\theta^{S/D}_P)$ denotes principal operators (parameters) corresponding to singles (S) or doubles (D) excitations, $N_S$ and $N_D$ are the number of singles and doubles excitation operators (parameters)
in the principal subspace respectively
and $\hat{U}_P$ denotes the principal unitary. The ordering of the operators in the ansatz is such that for doubles and singles unitary blocks
$\mid \theta^{D_1}_P \mid > \mid \theta^{D_2}_P \mid > ... > \mid \theta^{D_{N_D}}_P \mid$ and $\mid \theta^{S_1}_P \mid > \mid \theta^{S_2}_P \mid > ... > \mid \theta^{S_{N_S}}_P \mid$
conditions are respectively maintained.
The selected principal parameters are then optimized via VQE.
We call this selection and optimization protocol as screened MP2-VQE (sometimes referred to as MP2S-VQE in this work).
This selection protocol is especially advantageous in the presence of noise, as it does not depend on noisy quantum measurements and
thus the ordering of the operators in the ansatz is structually robust from a many-body perspective.
Contrarily, constructing adaptive-growth ansatz through quantum measurements can be vulnerable to 
the disruptive effects of noise and alter the ideal choice of the relevant many-electron operators\cite{robledo2024chemistry}.
Once the principal parameters are included into the ansatz and optimized, the operators from the entire pool
are included into the auxiliary space following the same notion as in section \ref{adapt for principal selection}.

\section{Computational Overhead and Theoretical Bounds for Auxiliary Subspace Corrections} \label{ASC measurement overhead}

Once the user-defined method $X$ is chosen (as discussed in section \ref{principal space selection}), the additional overhead for ASC calculations require
only the calculations of single and double commutator expectation values of the form
\begin{equation} \label{single commutator for ASC}
\bra{\Phi_P}[\hat{B}, \hat{\mathcal{G}}_{A_i}]\ket{\Phi_P}    
\end{equation}
% $\bra{\Phi_P}[\hat{B}, \hat{\mathcal{G}}_{A_i}]\ket{\Phi_P}$
and
\begin{equation} \label{double commutator for ASC}
    {\bra{\Phi_P}\Big[ [\hat{B}, \hat{\mathcal{G}}_{A_i}], \hat{\mathcal{G}}_{A_i} \Big]\ket{\Phi_P}}
\end{equation}
% ${\bra{\Phi_P}\Big[ [\hat{B}, \hat{\mathcal{G}}_{A_i}], \hat{\mathcal{G}}_{A_i} \Big]\ket{\Phi_P}}$
where $\hat{A}_i$ is $i$-th auxiliary index. These commutator terms are used for both principal-to-auxiliary mapping (Eq. \eqref{theta_A as function of theta_P}) and
additive energy correction terms (Eq. \eqref{cost function with ASC}).
Note that the depth of the circuits corresponding to these calculations are governed by $\ket{\Phi_P}$ which consists only principal
parameters. Thus ASC does not require additional quantum circuit resources such as CNOT gates. The maximum number of additional commutator expectation calculations
is bounded by $2N_A$, stemming from $N_A$ single-commutators (Eq. \eqref{single commutator for ASC}) and $N_A$ double commutators (Eq. \eqref{double commutator for ASC}), where $N_A$ is the number of total auxiliary parameters. If X=ADAPT-VQE, this additional overhead reduces to
$<2N_A$ expectations since
$\bra{\Phi_P}[\hat{B}, \hat{\mathcal{G}}_{A_i}]\ket{\Phi_P}$-type commutator expectations are just by-products of ADAPT-VQE macro-iteration cycles
and do not require to be calculated separately.
Thus, the additional overhead required for ASC involves only some extra commutator expectation evaluations and scales as $\mathcal{O}(N_A)$, which remains significantly lower than the total
number of function evaluations for typical VQE optimizations.
% So, the additional overhead for ASC requires just some commutator expectation calculations and scales as
% $\mathcal{O}(N_A)$ which is always much less than the total number of function evaluations required for VQE optimizations.

To summarize, the AD(X)-ASC consists of the following steps
\begin{enumerate}
    \item Define the principal and auxiliary subspaces using any adaptive-growth ansatz construction protocol \enquote{X}, like ADAPT-VQE
    or MP2 based selection as discussed in section \ref{adapt for principal selection} and \ref{mp2 for principal selection}:
    \begin{equation}
        \{\theta\} = \{ \theta_P \} + \{ \theta_A \} 
    \end{equation}

    \item Optimize the cost function defined only in the reduced dimensional principal subspace:
    \begin{equation}
        \arg\min_{\theta_P} f(\boldsymbol{\theta_P}) = \arg\min_{\theta_P} \bra{\Phi_0} \hat{U}(\boldsymbol{\theta_P})^{\dagger} \hat{B} \hat{U}(\boldsymbol{\theta_P}) \ket{\Phi_0}
    \end{equation}
    For ADAPT-VQE steps 1 and 2 run in alternate cycles until the criteria (Eq. \ref{adapt convergence criteria}) is satisfied. For the MP2 based selection approach
    steps 1 and 2 can be performed sequentially.

    \item With the optimized set of principal parameters, the auxiliary parameters are reconstructed via Eq. \eqref{theta_A as function of theta_P} as a one-step process.
    This can be processed in parallel.

    \item Auxiliary subspace corrections are added to the optimized cost-function via Eq. \eqref{cost function with ASC} for better accuracy with no additional hardware
    requirements.

\end{enumerate}

The ultimate goal of AD(X)-ASC is to select and optimize the most relevant subspace of the unitary which is as small as possible via the method \enquote{X} and subsequently incorporate the effects
of the missing part of the unitary into the cost-function via ASC.
In Section \ref{result and discussion} we will discuss the numerical benefits
of our method.

% \begin{figure*}[!ht]
%     \centering  
% \includegraphics[width=\textwidth]{ad_adapt_asc_comparison_plot.png}
% \caption{\textbf{Comparison between ADAPT-VQE and AD(ADAPT-VQE)-ASC for different molecular geometries:} The three columns are for linear $H_4$ chain, $BeH_2$ and $LiH$ molecule. Here y axes in all the subplots
% correspond to internuclear distances. The y-axes in (a), (b), (c) shows the energy difference from FCI and (d), (e), (f) corresponds to the associated number of CNOT gates. The solid and dotted lines correspond to
% ADAPT-VQE and AD(ADAPT-VQE)-ASC respectively with a specific threshold $\epsilon$. Note that AD(ADAPT-VQE)-ASC does not
% require additional CNOT gates in the circuit, so number of gates for it remains the same with ADAPT-VQE.}
%     \label{ad(adapt)-asc_CCSD(TQ)_pot_en_surf}
% \end{figure*}

\begin{figure*}[!ht]
    \centering  
\includegraphics[width=\textwidth]{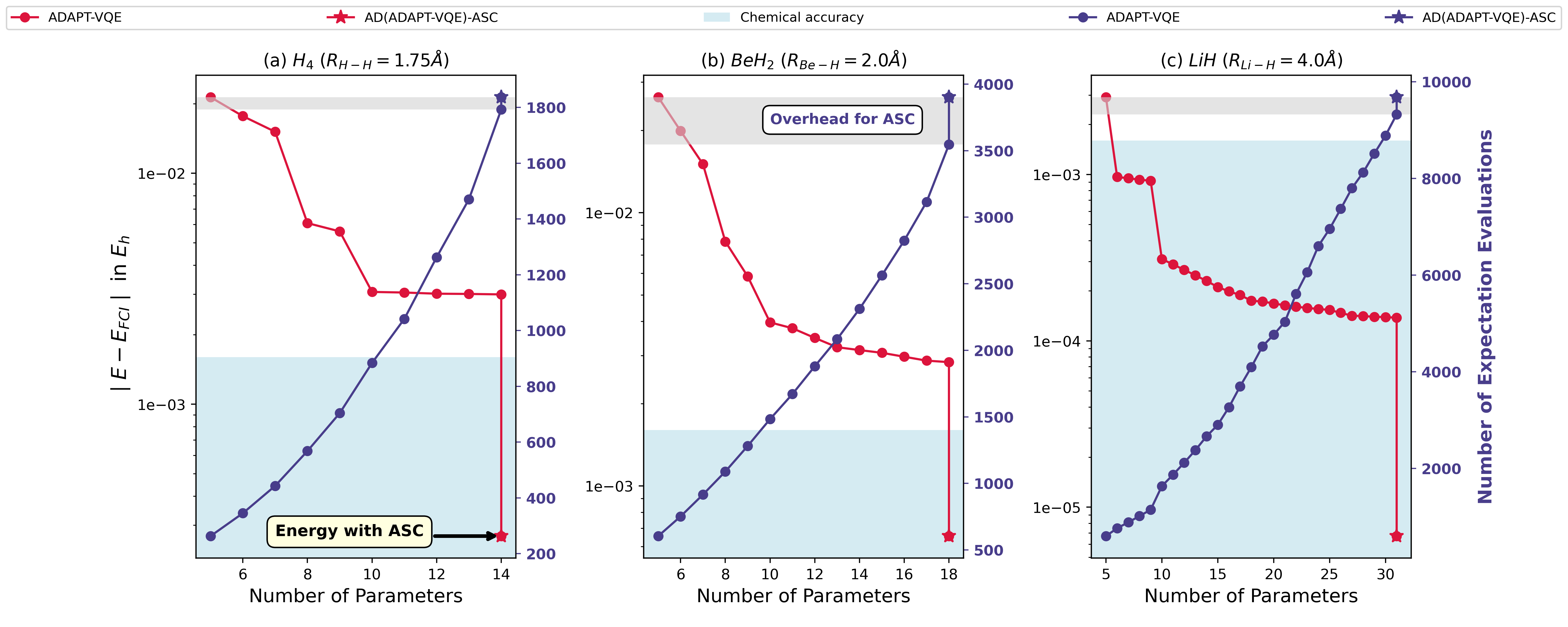}
\caption{\textbf{Energy optimization landscape for AD(ADAPT-VQE)-ASC with recycled initialization:} The x-axis shows the number of parameters in the ansatz and the left y-axis corresponds to the logarithm of the difference in energy with FCI (the global minimum)
for stretched geometries of (a) $H_4$ linear chain ($R_{H-H}=1.75\text{\AA}$), (b) $BeH_2$ ($R_{Be-H}=2.0\text{\AA}$) and (c) $LiH$ ($R_{Li-H}=4.0\text{\AA}$). The right y-axis
shows the number of expectation evaluations that includes both gradient calculations for operator selection and function evaluations for optimization. The initial energy
trajectory is for conventional ADAPT-VQE. Once the convergence criteria is met with the threshold $\epsilon=10^{-3}$, ADAPT-VQE exits the selection and optimization cycle and one-step ASC is performed at the last step. It shows the signature plummeting effect of ASC where energy reaches a more optimal minima for all the cases shown here. The corresponding measurement overhead for ASC is also shown in the grey shaded area along the right y-axis. The chemical accuracy is taken to be $1.6mE_h$ energy difference 
from FCI.
% without requiring any additional parameters into the ansatz. Due to the strong correlation effects, ADAPT-VQE energies for $H_4$ and $BeH_2$ lie outside the chemical accuracy window and ASC provides a one order-of-magnitude more accurate energy that lies well within the chemical accuracy.
}
    \label{ad(adapt)-asc_CCSD(TQ)_energy_landscape}
\end{figure*}

\section{Results and Discussion} \label{result and discussion}

In this section we will numerically demonstrate the superior performance and generailty of the theoretical framework developed so far
in electronic structure problems. We will consider two methods: (1) ADAPT-VQE and (2) screened MP2-VQE to select the principal
parameter space as discussed in section \ref{principal space selection}.
Our aim is to numerically study how effectively ASC can mitigate the effects of local minimas and provide more accurate
results without utilising any significantly additional quantum or classical overhead.
The results are obtained by in-house codes using python libraries and Qiskit\cite{qiskit2024}.
The electronic integrals are imported from
PySCF\cite{sun2018pyscf}.
In case of ADAPT-VQE,
we did not use the default
Qiskit implementation of the same and the threshold ($\epsilon$)
used here is not necessarily the same as that of the original paper\cite{grimsley2019adaptive}.
The operator selection is done manually using the calculation of commutators along with the threshold criterion
as described in Eq. \eqref{adapt gradient equation} and \eqref{adapt convergence criteria}.
The subsequent VQE optimization is done using VQE function from Qiskit Nature with maximum iteration (maxiter) set to be 10,000.
We have used L-BFGS-B and COBYLA optimizers for noiseless and noisy studies respectively.
% \textcolor{red}{NOT ALL THE CALCULATIONS HAVE THE ENTIRE PES. THUS SYMMETRIC STRETCHING IS CONFUSING...PERHAPS YOU DO NOT NEED TO MENTION THE SYSTEMS IN THIS SUBSECTION AND MENTION THEM ONLY WHEN THEY ARE DISCUSSED.}
% The numerical calculations were performed on three molecules with different electronic complexities
% - (a) $H_4$ linear chain, symmetric stretching of (b) $BeH_2$
% % \textcolor{red}{(SYMMETRIC STRETCHING OF H4 AND BEH2? LIH CANNOT HAVE SYMMETRIC STRETCHING AS IT HAS JUST ONE BOND)}
% and (c) $LiH$. 
For all the calculations STO-6G basis set is used.
% \textcolor{red}{(ALL THE CALCULATIONS USING STO-6G?)}
% if not mentioned explicitly otherwise.
Jordan-Wigner
transformation is applied for the fermionic-to-qubit mapping.
For all the simulations, the principal unitary was constructed 
from an operator pool consisting of singles and doubles (SD) excitations only. The auxiliary space contains the entire set of singles, doubles, triples and quadruples (SDTQ) operators. No symmetry considerations were used to reduce the qubit
requirements.
% For noisy studies CNOT-efficient quantum
% circuits\cite{yordanov2020efficient} were used.

The numerical discussion section is divided into three major sections. In Section \ref{x=adapt vqe}
we consider ADAPT-VQE as a tool for the principal subspace selection under noiseless scenarios.
Section \ref{screened mp2 vqe result} contains results with the screened MP2-VQE which is structurally more
robust under hardware noise than measurement-based ansatz construction protocols. With screened MP2-VQE
we discuss the effect of ASC under noisy circumstances as well. Finally in section \ref{initialization technique}
we introduce a generator-informed initialization strategy for ADAPT-VQE that can lead to faster convergence.
% mitigate local
% traps in their optimization landscape more effectively.

% the advantages of a novel initialization technique is discussed over recycled or Hartree-Fock
% initializations.

\begin{figure*}[!ht]
    \centering  
\includegraphics[width=\textwidth]{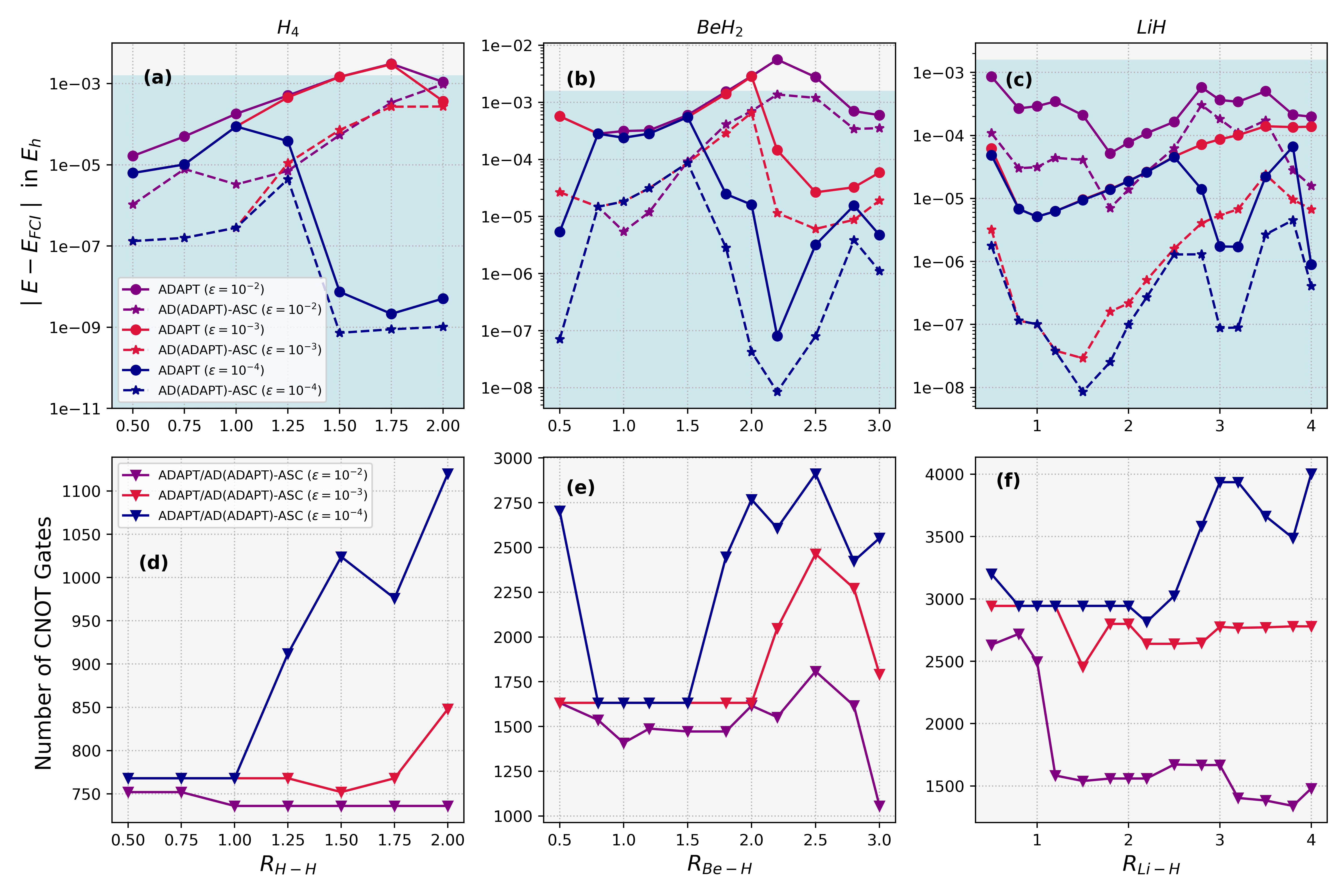}
\caption{\textbf{Comparison between ADAPT-VQE and AD(ADAPT-VQE)-ASC for different molecular geometries:} The three columns are for linear $H_4$ chain, $BeH_2$ and $LiH$ molecule. Here y axes in all the subplots
correspond to internuclear distances. The y-axes in (a), (b), (c) show the energy difference from FCI and (d), (e), (f) correspond to the associated number of CNOT gates. The solid and dotted lines correspond to
ADAPT-VQE and AD(ADAPT-VQE)-ASC respectively with a specific threshold $\epsilon$. Note that AD(ADAPT-VQE)-ASC does not
require additional CNOT gates in the circuit, so number of gates for it remains the same with ADAPT-VQE.}
    \label{ad(adapt)-asc_CCSD(TQ)_pot_en_surf}
\end{figure*}

\subsection{Principal Unitary Selection by X = ADAPT-VQE} \label{x=adapt vqe}

In this section we use ADAPT-VQE as a tool to select the principal unitary which caters to an iterative construction
of the ansatz to select important operators as discussed in details in section \ref{adapt for principal selection}.
First we study some specific molecular configurations to demonstrate how ASC can partially mitigate the issue of local traps.
Then we move over to the comparison between ADAPT-VQE and AD(ADAPT-VQE)-ASC for the
entire potential energy profile of these systems.

\subsubsection{Local Traps in Energy Optimization Landscape and ASC induced Plummeting Effect} \label{adapt landscape study}

% \textcolor{red}{THIS IS NOT STRICTLY A DEPICTIVE ENERGY LANDSCAPE OF OUR METHOD; BUT RATHER THIS REPRESENTS THE ENERGY LANDSCAPE OF ADAPT. OUR CONTRIBUTION HERE IS IN JUST DEMONSTRATING THE NOSE-DIVE. SO THERE IS NO POINT IN ELABORATING UPON THE BURROWING EFFECTS FOR ADAPT WHICH IS KNOWN. WE CAN JUST TANGENTIALLY MENTION HERE ABOUT BURROWING, BUT MORE STRESS SHOULD BE ON PLUMMETING. SECTION VI SHOULD HIGHLIGHT THE BURROWING EFFECT MORE ELABORATELY AS THE GRADIENT INFORMED INITIALIZATION OFTEN HELPS IN FINDING A LOWER-ENERGY "BURROWING CHANNEL" THAN A RECYCLED PARAMETER INITIALIZATION TRAJECTORY.}

Fig. \ref{ad(adapt)-asc_CCSD(TQ)_energy_landscape} illustrates some typical energy optimization landscapes of AD(ADAPT-VQE)-ASC method
for some molecular systems at their fixed internuclear geometries.
% To demonstrate energy landscape plummeting due to non-iterative ASC, 
% the principal subspace selection is done via ADAPT-VQE.
Here the plot
contains twin y-axis where the left y-axis represents the logarithm of energy difference of 
the methods under consideration in $E_h$ with respect to the global minimum 
obtained via full configuration interaction
(FCI) calculation. Along the horizontal axis, the incremental number of parameters in 
the circuit is plotted. The right y-axes correspond to the number of 
expectation evaluations which includes both the number
of gradient evaluations for operator selection and cost-function evaluations 
for the VQE optimizations.
The threshold for the selection of the primary subspace via ADAPT-VQE was set to be $\epsilon = 10^{-3}$
% the principal parameters subset is selected in this case via ADAPT-VQE with 
% operator selection threshold $\epsilon=10^{-3}$.
and each micro-iteration is initialized with recycled parameters.
% As outlined in section \ref{adapt for principal selection},
% Once the ADAPT-VQE operator selection and optimization is completed, 
% the auxiliary parameters are reconstructed via Eq. \eqref{theta_A as function of theta_P}. The final energy is obtained by plugging the principal and
% the reconstructed auxiliary parameters in the expression 
% Eq. \eqref{cost function with ASC}, with the necessary operator
% substitutions discussed in section \ref{adapt for principal selection}.

Panel (a) of Figure \ref{ad(adapt)-asc_CCSD(TQ)_energy_landscape} depicts 
the optimization landscape shown for linearly stretched $H_4$ chain 
($R_{H-H}=1.75$\AA) which reveals that ADAPT-VQE ($\epsilon=10^{-3}$) energy 
profile stagnates at a local minimum.
This is evident as the energy trajectory shows no visible improvement 
over the last four operator selection cycles although they have non-vanishing 
gradient. However, once the convergence condition over the energy gradient 
(Eq. \eqref{adapt convergence criteria}) is met,
the ASC is activated, significantly reducing the energy error from
$3 \times 10^{-3}E_h$ for ADAPT-VQE to $2 \times 10^{-4}E_h$ which is well within the 
chemical accuracy. This corresponds to an accuracy improvement of more than an order of magnitude relative to the FCI energy.
% In Fig. \ref{adspqe_CCSD(TQ)_energy_landscape} (a)
% for linear $H_4$ chain with stretched bonds, we can see ADAPT-VQE is sort of stuck in a local minima as in the
% last four operator selection cycles no visible improvement is observed in energy trajectory. However, once this
% cycle converges with the condition in Eq. \eqref{adapt convergence criteria}, the post-optimization auxiliary subspace correction gets activated and it brings the energy
% substantially down (from energy error $1.4 \times 10^{-3}E_h$ for ADAPT to $6.9 \times 10^{-5}E_h$ with ASC) to provide more than one order-of-magnitude better accuracy with respect to FCI energy.
% where $N_A$ is the number of auxiliary parameters stemming from the denominator of each predicted auxiliary parameters
% from Eq. \eqref{theta_A as function of theta_P}.
For $BeH_2$ (with $R_{Be-H}=2$\AA) (Fig. \ref{ad(adapt)-asc_CCSD(TQ)_energy_landscape}(b)),
ADAPT-VQE shows similar behavior
while ASC still provides much better accuracy ($\sim 6 \times 10^{-4} E_h$) compared to ADAPT-VQE ($\sim 3 \times 10^{-3} E_h$).
Similar trend is observed for
$LiH$ ($R_{Li-H}=4$\AA) where a local minima is still quite conspicuous toward the end of the ADAPT-VQE iterations.
% and ASC \textcolor{red}{provides more than one order of magnitude} accuracy ($\sim 9.5 \times 10^{-6} E_h$)
% over ADAPT-VQE ($\sim 1.35 \times 10^{-4}E_h$).
As analyzed previously, such improvements due to ASC requires no additional circuit resources such as
CNOT gates or extra qubits, and only requires an insignificant number of additional
gradient-like measurements (over the conventional choice of the method
X where X=ADAPT-VQE in this section).
% , and the extra measurements are depicted by the horizontal grey 
% shaded region}. 
Quantitatively, the number of extra gradient-like commutator
expectation values is bounded by $\leq 2N_A$ as discussed in section \ref{ASC measurement overhead}.
The grey shaded region along right y-axes in Fig. \ref{ad(adapt)-asc_CCSD(TQ)_energy_landscape}
shows the number of extra expectation values to be calculated which is
proportional to the measurement overhead required for performing ASC.
% For example, in case of $LiH$ the number of SDTQ operators in STO-6G basis are 224, out of which the total number of SDs is 92. In this case ASC requires only 356 extra expectation values to be calculated where
% 224 are of type Eq. \eqref{double commutator for ASC} for SDTQ and 132 are of type Eq. \eqref{single commutator for ASC} for TQ only since the other 92 single commutators for SD
% are a by-product of ADAPT-VQE iterations itself (see Eq.  \eqref{adapt gradient equation} \eqref{adapt convergence criteria}).
Evidently, this overhead (grey shaded region) is 
insignificant relative to total number of expectation evaluations.
% absolutely nominal with respect to the total number of expectation evaluation required for any VQE calculations.
Interestingly, for a particular molecule in a given basis set, this overhead is invariant with respect to different geometries or different ADAPT-VQE threshold.

\begin{figure*}[!ht]
    \centering  
\includegraphics[width=\textwidth]{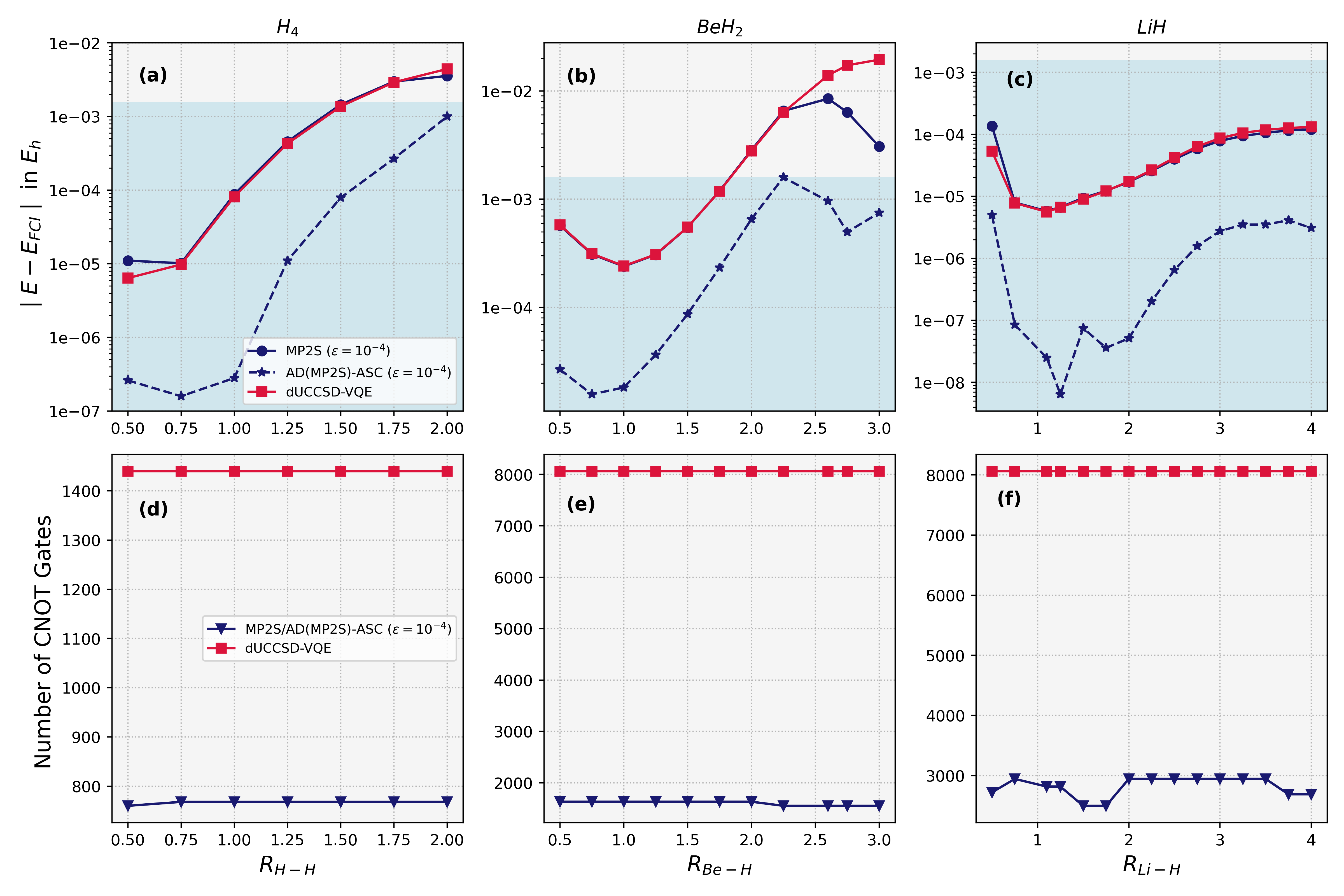}
\caption{\textbf{Comparison between MP2S-VQE and AD(MP2S-VQE)-ASC for different molecular geometries:} The axes information are same as Fig. \ref{ad(adapt)-asc_CCSD(TQ)_pot_en_surf} with MP2S-VQE performed with only one
threshold $\bar{\epsilon}=10^{-4}$. The dUCCSD energy and CNOT count values are also shown here for comparison. Note that the CNOT count for
AD(MP2S-VQE)-ASC is same as MP2S-VQE.}
    \label{ad(mp2)-asc_CCSD(TQ)_pot_en_surf}
\end{figure*}

In this context, it is worth noting that ADAPT-VQE can partially alleviate the challenge of local minima by progressively adding operators, enabling it to \enquote{burrow} into the parameter landscape \cite{Grimsley2023}. However, this effect is most pronounced for very small values of $\epsilon$ with relatively more number of parameters, which typically leads to deeper circuits
and excessively high measurement overhead. Our approach additionally achieves a one-step \enquote{plummeting effect}, reaching a significantly better optimal minimum even at higher $\epsilon$ values, all while maintaining the same number of parameters and circuit depth as that of ADAPT-VQE.
Thus a combination of ADAPT-VQE (at a higher $\epsilon$ value) with ASC leads to extremely accurate energy estimation
at much lower quantum resource requirements even with a set of 
recycled parameters, which can be significantly improved upon further with a 
generator-informed initialization (\textit{vide infra}).

% \begin{figure*}[!ht]
%     \centering  
% \includegraphics[width=\textwidth]{ad_adapt_asc_comparison_plot.png}
% \caption{\textbf{Comparison between ADAPT-VQE and AD(ADAPT-VQE)-ASC for different molecular geometries:} The three columns are for linear $H_4$ chain, $BeH_2$ and $LiH$ molecule. Here y axes in all the subplots
% correspond to internuclear distances. The y-axes in (a), (b), (c) shows the energy difference from FCI and (d), (e), (f) corresponds to the associated numbers of CNOT gates. The solid and dotted lines correspond to
% ADAPT-VQE and AD(ADAPT-VQE)-ASC respectively with a specific threshold $\epsilon$. Note that AD(ADAPT-VQE)-ASC does not
% require additional CNOT gates in the circuit, so number of gates for it remains the same with ADAPT-VQE.}
%     \label{ad(adapt)-asc_CCSD(TQ)_pot_en_surf}
% \end{figure*}

% \begin{figure*}[!ht]
%     \centering  
% \includegraphics[width=\textwidth]{ad_mp2_asc_comparison_plot.png}
% \caption{\textbf{Comparison between screened MP2-VQE and AD(MP2-VQE)-ASC for different molecular geometries:} The axes informations are same as Fig. \ref{ad(adapt)-asc_CCSD(TQ)_pot_en_surf} with screened MP2-VQE performed with only one
% threshold $\bar{\epsilon}=10^{-4}$. The dUCCSD energy and CNOT count values are also shown here for comparison.}
%     \label{ad(mp2)-asc_CCSD(TQ)_pot_en_surf}
% \end{figure*}

\subsubsection{Accuracy Over the Potential Energy Surface: ADAPT-VQE vs AD(ADAPT-VQE)-ASC}

% \textcolor{red}{Introductory sentences needed stating that in this section, 
% we shall be studying 
% the PES with two different choices of X and irrespective of X, ASC leads to an 
% improved energetic description over X, over the entire PES. }

% \subsubsection*{X= ADAPT-VQE}

In Fig. \ref{ad(adapt)-asc_CCSD(TQ)_pot_en_surf} we have studied the potential energy surfaces (PES) for all the molecules under consideration
to assess the comparative behavior of ADAPT-VQE and AD(ADAPT-VQE)-ASC for different electronic complexities.
Here we have considered ADAPT-VQE to select the principal subset with three different thresholds $\epsilon = 10^{-2}, 10^{-3}$ and $10^{-4}$.
In Fig. \ref{ad(adapt)-asc_CCSD(TQ)_pot_en_surf} ((a), (b), (c)) the logarithm of energy difference with respect to FCI (in $E_h$) is plotted in the y-axis
while the x-axis represents different molecular bond distances in \AA. Fig. \ref{ad(adapt)-asc_CCSD(TQ)_pot_en_surf} ((d), (e), (f))
% and ((g), (h), (i))
represent the CNOT count
% and number of parameters
for all different molecular geometries as a measure of quantum resource utilization.
The solid lines and dashed lines represent ADAPT-VQE and AD(ADAPT-VQE)-ASC with purple, red and blue indicating $\epsilon$ values $10^{-2}$, $10^{-3}$ and $10^{-4}$
respectively. 
In Fig. \ref{ad(adapt)-asc_CCSD(TQ)_pot_en_surf} (a) for linear $H_4$ it is observed that AD(ADAPT-VQE)-ASC outperforms ADAPT-VQE in both weak and strong
electronic correlation regime
while providing one to two orders-of-magnitude better energy accuracy. The effect of ASC is so prominent that for near equilibrium geometries of linear $H_4$ chain
(from $R_{H-H}=0.5$ to $R_{H-H}=1.25$\AA), our method with $\epsilon=10^{-2}$ consistently provides visibly better accuracy than even ADAPT-VQE with $\epsilon=10^{-4}$.
% At $R_{H-H}=1.25$\AA 
Similar trend is observed for symmetric bond stretching of $BeH_2$ as well where AD(ADAPT-VQE)-ASC always remains within the
chemical accuracy window even if ADAPT-VQE struggles to achieve the same with $\epsilon=10^{-2}$ and $10^{-3}$ at around $R_{Be-H}=2.0\AA$.
For $LiH$, AD(ADAPT-VQE)-ASC with $\epsilon=10^{-3}$ shows drastically more accurate results than ADAPT-VQE with $\epsilon=10^{-4}$
for both near equilibrium and stretched geometries (till $R_{Li-H}=2.8$\AA). At $R_{Li-H}=3.8$\AA, AD(ADAPT-VQE)-ASC with $\epsilon=10^{-2}$ provides $40\mu E_h$ better
energy accuracy (dotted purple line) with respect to ADAPT-VQE with $\epsilon = 10^{-4}$ (solid blue line) resulting in 61\% reduction in CNOT gates.
% For all the numerical studies, our method always stays well within the chemical accuracy even for the cases when ADAPT-VQE falls out of the
% chemical accuracy window.
As theoretically argued and depicted in Fig. \ref{ad(adapt)-asc_CCSD(TQ)_pot_en_surf} (d), (e), (f),  ADAPT-VQE and AD(ADAPT-VQE)-ASC both 
require same number of CNOT gates.

\subsection{Principal Unitary Selection by X = Screened MP2-VQE (MP2S-VQE)} \label{screened mp2 vqe result}
% \subsubsection*{X= Screened MP2-VQE} \label{mp2s vqe results}

In this case we select the principal subspace via MP2 values and spin symmetry as discussed in section \ref{mp2 for principal selection}.
Here the potential energy surface is studied with noiseless simulations followed by noisy simulations to demonstrate the signature
plummeting effect even under cases with realistic depolarising noise channels.

% \begin{figure*}[!ht]
%     \centering  
% \includegraphics[width=\textwidth]{ad_mp2_asc_comparison_plot.png}
% \caption{\textbf{Comparison between MP2S-VQE and AD(MP2S-VQE)-ASC for different molecular geometries:} The axes informations are same as Fig. \ref{ad(adapt)-asc_CCSD(TQ)_pot_en_surf} with MP2S-VQE performed with only one
% threshold $\bar{\epsilon}=10^{-4}$. The dUCCSD energy and CNOT count values are also shown here for comparison.}
%     \label{ad(mp2)-asc_CCSD(TQ)_pot_en_surf}
% \end{figure*}

\subsubsection{Potential Energy Surface Study} \label{mp2s vqe results}
In Fig. \ref{ad(mp2)-asc_CCSD(TQ)_pot_en_surf} we have shown the case with X= screened MP2-VQE (MP2S-VQE). Here the principal subspace is formed by MP2 screened doubles with
a threshold $\bar{\epsilon}= 10^{-4}$ along with single excitations
allowed by orbital-symmetry as discussed in section \ref{mp2 for principal selection}. Results for disentangled UCC with singles and doubles (dUCCSD-VQE) are also shown
for comparison with the default operator ordering as implemented in Qiskit. 
Since MP2S-VQE consists of dominant operators from SD pool, the results are identical with dUCCSD in most of the cases.
However, for $BeH_2$ (plot (b) in Fig. \ref{ad(mp2)-asc_CCSD(TQ)_pot_en_surf}) one can see in the stretched geometries (strong
correlation region), MP2S-VQE and dUCCSD-VQE deviates slightly. This is most likely due to different operator ordering,
dUCCSD-VQE gets stuck in a higher-energy local minima compared to MP2S-VQE. For all molecular cases studied here AD(MP2S-VQE)-ASC
provides chemically accurate results even in the strong correlation regions where MP2S-VQE and dUCCSD-VQE suffer
in terms of accuracy. For example, in the strongly correlated region for $BeH_2$ with $R_{Be-H}=2.75\text{\AA}$,
MP2S-VQE and dUCCSD-VQE energy errors from FCI are $0.0064 E_h$ and $0.017E_h$ respectively whereas, due to ASC
this difference reduces to $0.0005E_h$. Similar trend is observed for linear $H_4$ and $LiH$ throughout the potential energy
surface.
In this case the parallelity of the errors over the PES for MP2S-VQE and AD(MP2S-VQE)-ASC is capable of providing consistent energy shift across PES.

To summarize, the numerical studies in this section suggests that AD(X)-ASC provides one to two orders of magnitude better accuracy
than its conventional counterparts (i.e. method X) at a particular threshold even in the regions of molecular strong correlations
where method X suffers to account for chemically accurate energies. This enhanced accuracy is obtained at a nominal overhead with $2N_A$ number of extra commutator measurements
at worst and no additional circuit resources are required.
Since the numerical simulations are noiseless in this section,
in the next section we will discuss how quantum hardware noise affects such calculations.

\subsubsection{Plummeting Effect under Noisy Simulations}
\begin{figure}%[!ht]
    \centering
\includegraphics[width=\linewidth]{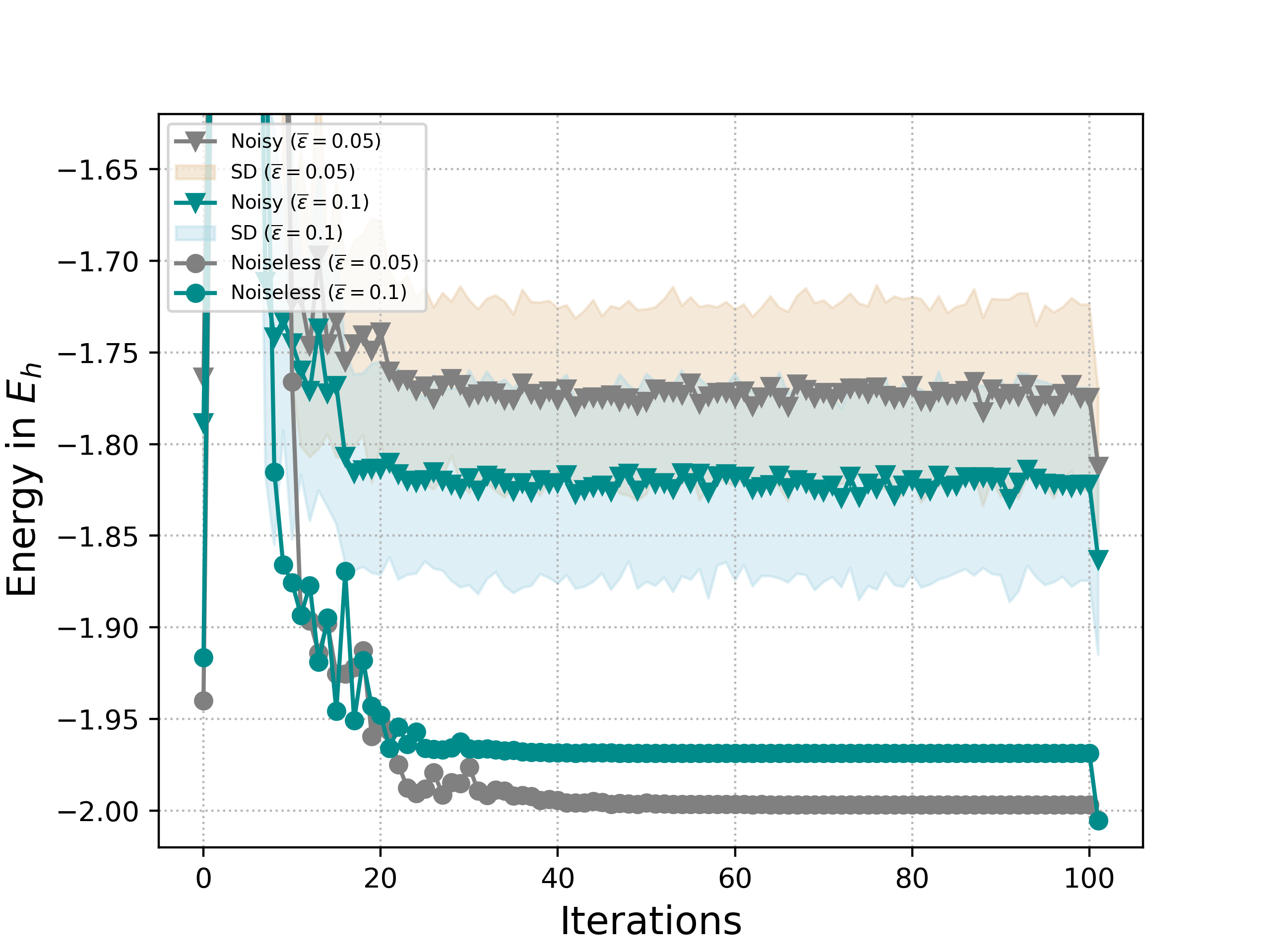}
\caption{\textbf{Noisy simulation:} Simulation with depolarising noise channel for $H_4$ at $1.5\text{\AA}$ with principal space selection done by MP2 screening at different thresholds $\bar{\epsilon}$. Energy in $E_h$ is plotted in y-axis and x-axis corresponds to number of iterations. The corresponding noiseless simulations are also shown for better comprehension of the disruptive effects of noise.
}
    \label{ad_adapt_asc_noise_plot}
\end{figure}

For implementations in noisy hardware, as measurement based operator selection is unreliable due to the error-prone expectation value
measurements, we opt for the MP2 based operator selection as discussed in section \ref{mp2 for principal selection}.
In Fig. \ref{ad_adapt_asc_noise_plot} we have shown both noisy and noiseless simulations with MP2 based operator selection.
The noisy simulations are done using only depolarising noise channels in Qiskit with two qubit gate error 0.01 and one qubit gate error 0.001
for linear $H_4$ chain with $H-H$ bond distance $1.5$\AA. We have used CNOT-efficient circuits\cite{yordanov2020efficient,magoulas2023cnot} for this study to reduce number of CNOT gates.
In addition to this, we have used Zero Noise Extrapolation\cite{giurgica2020digital} (ZNE) error mitigation technique. 
The value of the threshold ($\overline{\epsilon}$) regarding MP2 based selection is taken to be 0.05 (grey curve) and 0.1 (green curve) which selects
8 and 6 parameters respectively out of the 26 SD operators in the pool.  
Since in noisy simulations there is no notion of convergence, we terminated the simulations with 100 iterations
for both noiseless and noisy studies to treat all the cases on an equal footing. The noisy plots are obtained by averaging over 100 independent runs
and the corresponding standard deviations are shown with the shaded regions. Even for the noisy simulations, %with $\overline{\epsilon}=0.05$ and $0.1$,
the signature dip in average energy as well as in the corresponding standard deviation due to ASC is discernible. It is important to note 
that the average energy for $\overline{\epsilon}=0.1$ is better than
$\overline{\epsilon}=0.05$ under noise, which is counterintuitive as the former case has fewer parameters in the ansatz and hence ideally should be less accurate 
(as can be seen from the noiseless simulations in Fig. \ref{ad_adapt_asc_noise_plot}). This shows that the higher number of CNOT present in the circuit corresponding 
to $\overline{\epsilon}=0.05$ drives it away from the optimal solution 
due to the more disruptive signature of noise than the ansatz with
$\overline{\epsilon}=0.1$. In such a case, it is still evident from 
Fig. \ref{ad_adapt_asc_noise_plot} that the average energy (along with the standard deviation) with ASC for $\overline{\epsilon}=0.05$
is of the same order of the energy for $\overline{\epsilon}=0.1$ which shows the pronounced impact of ASC even under noisy scenarios.

\begin{figure*}[!ht]
    \centering  
\includegraphics[width=\textwidth]{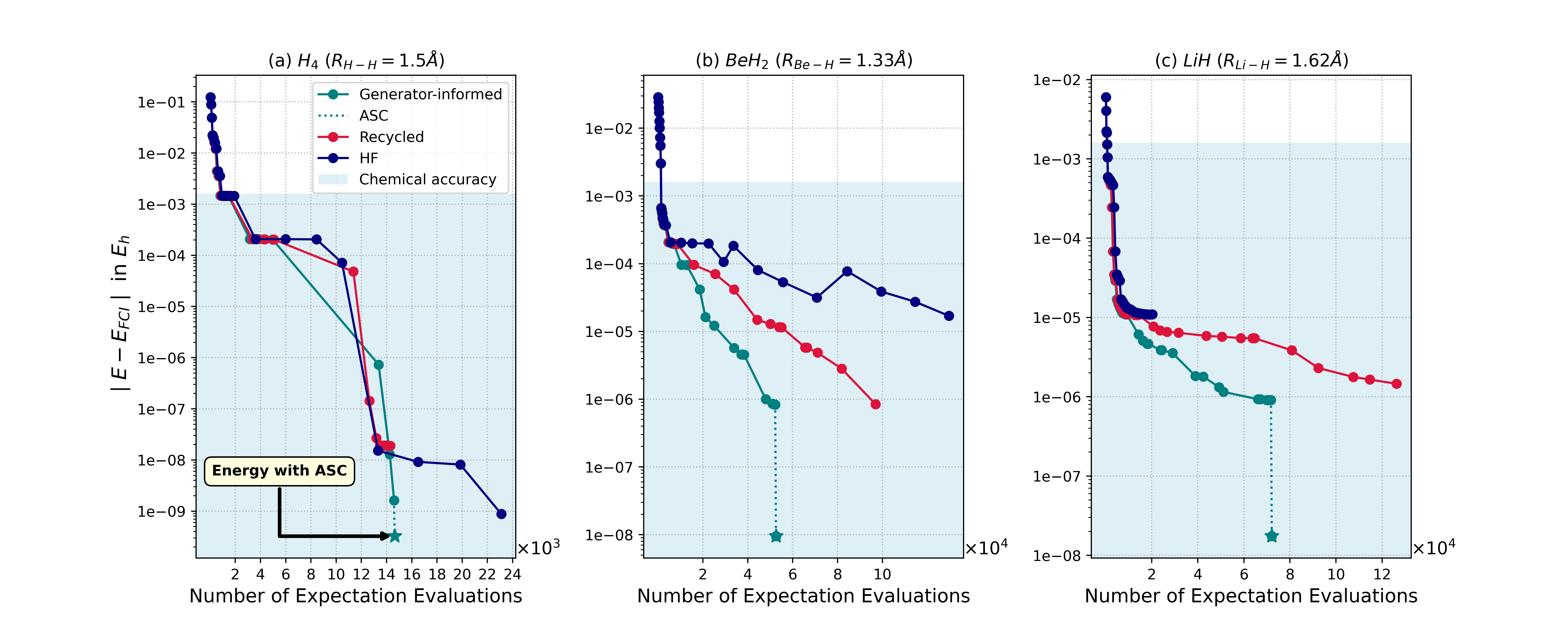}
\caption{\textbf{Comparison between generator-informed, recycled and HF initialization for ADAPT-VQE:} Energy error (in logarithm scale) for ADAPT-VQE ($\epsilon=10^{-5}$) with respect to FCI is plotted as a function of the number of cost-function evaluations for three different molecules. The numerical studies suggest generator-informed initialization converges much faster compared to HF or recycled initializations. For both $BeH_2$ and $LiH$, the generator-informed initialization shows almost 50\% reduction in number of function evaluations compared to recycled initialization. The ASC (dotted part of the green curve) leads to better minima with almost no additional
quantum resources.
% Note that non-iterative ASC is not included in this study here to demonstrate the advantage of generator-informed initialization strategy only.
}
    \label{mapped initialization plot}
\end{figure*}

\section{Generator-Informed Initialization Strategy for Improved ADAPT-VQE Optimization} \label{initialization technique}

Effective parameter initialization is one of the most crucial aspects for the trainability of PQCs and convergence of VQE calculations\cite{zhou2020quantum, larocca2025barren,cerezo2021variational}.
%since random initializations often get stuck in
%far-from-optimal local minimas. Often, randomly initialized deep unstructured PQCs lead to Barren Plateaus.
In cases where VQE cost-function landscapes flatten exponentially fast with the system size, initializing the
optimization with random values of the parameters often get stuck in local traps or barren plateaus (BP)\cite{mcclean2018barren, cerezo2021variational}.
Gradient descent-like algorithms can converge to any local minimum, mostly governed by the initialization\cite{bittel2021training, wierichs2020avoiding}.
In general, if the initial guess of the parameters lies within the same local trap as that of the global minimum, then the classical
optimizers seamlessly find the minimum. However, if this is not the case, then BPs obstruct the optimizer to escape the narrow gorge
and it results in a sub-optimal solution\cite{mao2024towards}.
Some recent studies show that parameter trainability can be enhanced via certain initialization techniques\cite{wang2024trainability}.
Thus, clever initialization strategies are one of the most promising ways to avoid BPs or local minimas since it jump start the training in a region of large gradients\cite{larocca2025barren}.

For molecular VQEs with UCC-like ansatze, the usual practice is to start with the physically motivated Hartree-Fock (HF) initializations where all the parameters start from zero
(which corresponds to the HF solution) and the optimizations reach a nearly optimal minima in most cases.
Grimsley et al. have recently suggested a \enquote{recycled} parameter initialization\cite{Grimsley2023} strategy
in ADAPT-VQE protocol where each VQE optimization starts with all the optimized values of the parameters in previous step along with the parameter corresponding to
the new selected operator starting from zero (discussed in Section \ref{adapt for principal selection}). Such a recycled initialization was shown to perform better than
HF initialization in some molecular-VQE applications.

Here we propose a new initialization strategy that stems from the notion of adiabatic decoupling.
For a particular iteration, ADAPT-VQE selects the parameter with the largest gradient.
% The principal parameter space may be conceived to be formed by all the operators already included till ($k-1$)-th iteration and
% the operator selected at the $k$-th step is considered to be auxiliary parameter whose
% initial guess is obtained via
% Since ADAPT-VQE appends the operators one at a time,
% for each iteration, there is still an implicit temporal hierarchy among the selected (principal) operators.
Following the principles of adiabatic decoupling and mapping functional (Eq. \eqref{theta_A as function of theta_P}),
at the $k$-th iteration, the newly added parameter can be approximated as a function of the previously optimized $(k-1)$ parameters. This yields an initial guess of the $k$-th parameter
given by:
\begin{equation} \label{generator informed initialization}
    \theta_{P_k}^{init}  \leftarrow - \frac{\bra{\Phi_P^{k-1}}[\hat{B}, \hat{\mathcal{G}}_{P_k}]\ket{\Phi_P^{k-1}}}{\bra{\Phi_P^{k-1}}\Big[ [\hat{B}, \hat{\mathcal{G}}_{P_k}], \hat{\mathcal{G}}_{P_k} \Big]\ket{\Phi_P^{k-1}}}
\end{equation}
where, $\ket{\Phi_P^{k-1}}$ is the ansatz formed at the end of the $(k-1)$-th macro-iteration step
\begin{equation}
    \ket{\Phi_P^{k-1}} = \hat{U}_{P_1}^{\dagger}(\theta_{P_1}^{opt})..\hat{U}_{P_{k-1}}^{\dagger}(\theta_{P_{k-1}}^{opt}) \ket{\Phi_0}.
\end{equation}
with $[\theta_{P_1}^{opt},...\theta_{P_{k-1}}^{opt}]$ being the set of optimized parameters
from $(k-1)$-th iteration.
% Note that at the initial step with $k=1$ and no principal parameters in the ansatz,
% the parameter starts with 
One may notice that Eq. \eqref{generator informed initialization} is structurally same as the principal-to-auxiliary mapping function
(Eq. \eqref{theta_A as function of theta_P}).
This implies that at the $k^{th}$ iteration, all the operators chosen
till $(k-1)^{th}$ step is considered to span the principal parameter space and the 
$k^{th}$ operator chosen at the $k^{th}$ step is conceived as 
an auxiliary 
parameter which is subsequently added to the ansatz and 
it is initialized via Eq. \eqref{generator informed initialization} for further optimization.
% Here, the principal parameter space may be conceived to be formed by all the operators already included till ($k-1$)-th iteration and
% the operator selected at the $k$-th step is picked up from the auxiliary pool at that step
As the parameter initialization
is driven by the corresponding generators, it will be referred to as
% \enquote{mapped}
\enquote{\textit{generator-informed}} initialization.
In the case of ADAPT-VQE with the new generator-informed initialization strategy, 
we warm-start each micro-iteration with the entire parameter set by the ($k-1$)
optimized parameters from the previous step of the iteration along
with the $k$-th parameter obtained via Eq. \eqref{generator informed initialization}.

Though in this paper we discuss such generator-informed initialization strategy in the context of ADAPT-VQE only, the structure
of Eq. \eqref{generator informed initialization} suggest that it is problem-independent and thus we hope it can be applicable to other optimization problems as well.
We will explore more about such initialization strategies in forthcoming publications.

%\hat{U}_{1}^{\dagger}(\theta_{1}^{opt})..\hat{U}_{k-1}^{\dagger}(\theta_{k-1}^{opt})

\subsubsection*{Emergence of Lower Energy Landscape Burrowing with Generator Informed Initialization}
In Fig. \ref{mapped initialization plot} we have shown the energy convergence for ADAPT-VQE with $\epsilon=10^{-5}$ as a function of the number of expectation evaluations
with three different initialization techniques: (i) generator-informed (green curve), (ii) recycled (red curve) and (iii) HF initialization
(blue curve) as discussed in section \ref{initialization technique}.
The numerical studies for (a) $H_4$ ($R_{H-H}=1.5$\AA), (b) $LiH$ ($R_{Li-H}=1.33$\AA) and (c) $BeH_2$ ($R_{Be-H}=1.62$\AA) show
that during the initial ADAPT-VQE macro-iteration cycles, all three types of parameter initialization techniques
converge nearly to the same minima. However, the advantage of 
generator-informed initialization is more discernible as the number of parameters 
in the circuit increases since this amounts to a higher dimensional principal parameter
subspace that are mapped to reconstruct auxiliary parameters.
% \textcolor{red}{ADD HF and random init}
Consequently, one can see that for $BeH_2$, ADAPT-VQE with 
generator-informed initialization shows a 48\% reduction in the overall number 
of expectation evaluations (which includes function evaluations and gradient calculations)
% takes much less (almost half) number of function evaluations
to converge to similar energy minima compared to recycled initialization.
Further, ASC provides almost two orders of magnitude more accurate 
energy minima (shown with an asterisk mark) at the cost of negligible 
amount of additional expectation calculations.
For $H_4$, although the effect of generator-informed initialization is not
markedly visible, the generator-informed initialization converges to slightly better
minima. For $LiH$, generator-informed initialization reaches an 
energetically better minima with only half the amount of expectation 
evaluations required than other two initialization techniques.

In general, with generator-informed initialization for AD(ADAPT)-ASC,
one obtains two-fold advantages: firstly, it is heuristically observed that the optimization
trajectory finds a lower-energy \enquote{burrowing channel} that can potentially 
converge to lower energy minima compared to recycled or HF parameter initialization
with significantly reduced quantum measurements. Secondly, the ASC leads to
additional plummeting in the energy landscape without requiring any further 
quantum hardware resources, making the overall energy estimation extremely accurate.

\section{Conclusion and Future Outlook}

In this work, we introduce a formalism aimed at
optimizing the trade-off between resource efficiency and accuracy in parameterized quantum circuits. Our approach conjectures the presence of a 
temporal hierarchy of the parameter optimization landscape which natuarally 
leads to the decoupling of
%is based upon the
%conjecture that the multi-variable nonlinear optimization problems like VQE involve parameters that evolve towards a fixed point in vastly
%different timescales.
%Considering such temporal hierarchy, one can decouple 
the entire operator (parameter) space into a dominant
principal and a recessive auxiliary subspace.
Based upon this decoupling, the adiabatic approximation can be invoked by freezing the successive variation of the auxiliary parameters.
Starting from the parameter-shift rule, adiabatic approximation leads to a principal-to-auxiliary mapping functional that enables the reconstruction of auxiliary parameters from optimized principal parameters
only, leading to one-step auxiliary subspace corrections (ASC).
% the entire parameter space can be decoupled into a faster converging damped auxiliary parameters
% and a set of principal parameters which take much more number of iterations to converge. This decoupling leads to the adiabatic approximation
% via which the iterative variation of auxiliary parameters can be suppressed. This eliminates the coupling between principal and auxiliary parameters,
% confining the optimization to the reduced-dimensional principal subspace only. 
% We established a principal-to-auxiliary mapping that enables the reconstruction of auxiliary parameters from optimized principal parameters, leading to one-step auxiliary subspace corrections (ASC).
This correction mechanism, implemented without introducing additional parameters into the PQC, ensures that circuit depth remains constrained to the principal subspace.
Thus the correction terms factor in the approximate effects of the entire operator pool into the cost function towards a more optimal minima,
without incurring any additional quantum resource overheads or classical optimization
cycles which makes it an extremely resource efficient algorithm.
% , thus preserving resource efficiency.
Our AD(X)-ASC method, where \enquote{X} being the method to select and optimize the principal
parameter subspace, is a general framework that can be applied to a wide class of optimization tasks.

In this paper, we have shown two different possible methods to choose \enquote{X}, namely
ADAPT-VQE and MP2S-VQE
to treat challenging electronic correlation effects for molecular cases like bond stretching in
linear $H_4$ chain, $BeH_2$ and $LiH$.
% We have argued that the state-of-the-art ADAPT-VQE follows this adiabatic principle as well and it can be used as a tool to select the principal subspace.
% To avoid the measurment overhead in ADAPT-VQE for operator selection, we have also proposed measurement-free approaches based on MP2 screening techniques
% as a principal subspace selection protocol.
The numerical demonstrations show AD(ADAPT-VQE)-ASC and AD(MP2S-VQE)-ASC methods can provide upto two orders-of-magnitude
better energy accuracy than both ADAPT-VQE and MP2S-VQE respectively, without any additional optimization or quantum circuit resources.
The energy landscape study shows heuristic evidences that ASC exhibits a \enquote{plummeting} effect by which it can effectively alleviate local traps and reach a more optimal minima
without requiring additional parameters in the PQC.
The noisy simulations
suggest that the signature plummeting effect in the energy landscape exists even under the disruptive
behavior of noise which can be beneficial under the NISQ era of quantum computing.
Additionally we have proposed a generator-informed initialization technique which is entirely guided by the generator of
unitary rotations of the ansatz. Our numerical heuristics suggest this generator-informed initialization combined with ASC can lead to faster convergences 
by finding lower energy burrowing channels in the optimization landscape, followed by the characteristic ASC-aided plummeting for a more optimal energy minima estimation
with less quantum measurement requirements
% as well as more optimal energy minima estimation
compared to the conventional recycled
or HF based parameter initialization techniques.
From a general perspective, the structure of AD(X)-ASC shows promise that the optimization can be less affected by hardware noise
as the optimization task is projected on to a lower dimensional manifold.
Simultaneously, the problem of local traps, prevalent in shallow underparametrized circuits, is mitigated via generator-informed initialization along with ASC.
However, this needs further investigation which will be the subject of our future publications.

\section{Acknowledgments}
The authors acknowledge Ms. Dipanjali Halder for stimulating discussions.
RM acknowledges the 
financial support from Industrial Research and
Consultancy Centre (IRCC), IIT Bombay and 
Science and Engineering Research Board (SERB), Government
of India (Grant Number: MTR/2023/001306).
CP acknowledges University Grants Commission (UGC) for the fellowship.

\section*{AUTHOR DECLARATIONS}
\subsection*{Conflict of Interest:}
The authors have no conflict of interest to disclose.

\section*{Data Availability}
The data is available upon reasonable request to the corresponding author.

\appendix

\section{Second Order Approximation of the Cost Function}
\label{cost function approx appendix}

% \begin{equation} \label{f_theta expansion}
% \begin{split}
%     &f(\boldsymbol{\theta} \pm \frac{\pi}{2}\hat{e}_{A_i}) = \bra{\Phi_0}\prod_{P_I} \hat{U}_{P_I}^{\dagger} . . . \hat{U}_{A_i}^{\dagger}(\theta_{A_i} \pm \frac{\pi}{2}). . . \\
%     & \hat{U}_{A_2}^{\dagger} \hat{U}_{A_1}^{\dagger} \cdot \hat{B} \cdot \hat{U}_{A_1} \hat{U}_{A_2} ... \hat{U}_{A_i}(\theta_{A_i} \pm \frac{\pi}{2}) ... \prod_P \hat{U}_P \ket{\Phi_0}
% \end{split}
% \end{equation}
% Here, the auxiliary part of the unitary is expanded to explicitly reflect the parameter shift in the
% $A_i$-th component.

Starting from Eq. \eqref{f_theta expansion}, we can explicitly apply BCH expansion for auxiliary part of the unitary
% By employing the Baker-Campbell-Hausdorff (BCH) expansion for auxiliary unitaries, terms up to second order in $\theta_{A_i}$
% are retained, resulting in
\begin{equation}\label{f_theta_plus_minus_pi_by_2_upto_second_order_appendix}
\begin{split}
    & f(\boldsymbol{\theta}\pm\frac{\pi}{2}\hat{e}_{A_i}) = \bra{\Phi_P} \hat{B} \ket{\Phi_P} + (\theta_{A_i}\pm\frac{\pi}{2}) \\
    & \bra{\Phi_P}[\hat{B}, \hat{\mathcal{G}}_{A_i}]\ket{\Phi_P} + \sum_{A_{j \neq i}} \theta_{A_j} \\
    &  \bra{\Phi_P}[\hat{B}, \hat{\mathcal{G}}_{A_j}]\ket{\Phi_P} + \frac{1}{2} (\theta_{A_i}\pm\frac{\pi}{2})^2 \\
    & \bra{\Phi_P}\Big[ [\hat{B}, \hat{\mathcal{G}}_{A_i}], \hat{\mathcal{G}}_{A_i} \Big] \ket{\Phi_P} 
    + \frac{1}{2} \sum_{A_j} \sum_{A_k\ne A_i} (\theta_{A_j}\pm\frac{\pi}{2}\delta_{A_iA_j}) \theta_{A_k} \\
    &\bra{\Phi_P}\Big[ [\hat{B}, \hat{\mathcal{G}}_{A_j}], \hat{\mathcal{G}}_{A_k} \Big] \ket{\Phi_P} + . . .
    % + \sum_{A_j, A_k} \theta_{A_j} \theta_{A_k} \bra{\Phi_P}\Big[ [\hat{B}, \hat{\mathcal{G}}_{A_j}], \hat{\mathcal{G}}_{A_k} \Big] \ket{\Phi_P}
\end{split}    
\end{equation}
which is a non-terminating series. Here the Kronecker-Delta function $\delta_{A_iA_j}$ ensures that the parameter-shift of $\frac{\pi}{2}$ only
exists when $A_j$ is equal to $A_i$. Numerically the \enquote{non-diagonal} double commutators involving $\hat{\mathcal{G}}_{A_j}$
and $\hat{\mathcal{G}}_{A_k}$ along with the higher order terms in the series have negligible numerical contributions compared to the other leading order terms.
Thus they can be ignored which leads to the approximated Eq. \eqref{f_theta_plus_minus_pi_by_2_upto_second_order}.

\section{Principal-to-Auxiliary Mapping from Parameter-Shift Rule} \label{principal to auxiliary mapping from params shift rule appendix}

One can directly plug the full BCH expansion in Eq. \eqref{f_theta_plus_minus_pi_by_2_upto_second_order_appendix} into
the Parameter-shift rule for auxiliary gradient calculations (Eq. \eqref{delta theta_A})

\begin{equation}
    \begin{split} \label{delta theta_A expanded full}
        & \Delta \theta_{A_i} = \frac{1}{2} \Big[f(\boldsymbol{\theta}+\frac{\pi}{2}\hat{e}_{A_i}) - f(\boldsymbol{\theta}-\frac{\pi}{2}\hat{e}_{A_i})\Big]\\
        & = \frac{1}{2} \Big[ \Big(  \frac{\pi}{2} \bra{\Phi_P}[\hat{B}, \hat{\mathcal{G}}_{A_i}]\ket{\Phi_P} + \theta_{A_i}.\frac{\pi}{2} \\
        & \bra{\Phi_P}\Big[ [\hat{B}, \hat{\mathcal{G}}_{A_i}], \hat{\mathcal{G}}_{A_i} \Big] \ket{\Phi_P} +
        \frac{1}{2} \sum_{A_k \ne A_i} (\theta_{A_i}+\frac{\pi}{2})\theta_{A_k} \\
        & \bra{\Phi_P}\Big[ [\hat{B}, \hat{\mathcal{G}}_{A_i}], \hat{\mathcal{G}}_{A_k} \Big] \ket{\Phi_P} +... \Big) - \Big( -\frac{\pi}{2} \bra{\Phi_P}[\hat{B}, \hat{\mathcal{G}}_{A_i}]\ket{\Phi_P} \\
        & - \theta_{A_i}.\frac{\pi}{2} \bra{\Phi_P}\Big[ [\hat{B}, \hat{\mathcal{G}}_{A_i}], \hat{\mathcal{G}}_{A_i} \Big] \ket{\Phi_P} \\
        & + \frac{1}{2} \sum_{A_k \ne A_i} (\theta_{A_i}-\frac{\pi}{2})\theta_{A_k} \bra{\Phi_P}\Big[ [\hat{B}, \hat{\mathcal{G}}_{A_i}], \hat{\mathcal{G}}_{A_k} \Big] \ket{\Phi_P} +... \Big) \Big]
    \end{split}
\end{equation}
where, all the other terms are identical in both $f(\boldsymbol{\theta}+\frac{\pi}{2}\hat{e}_{A_i})$ and $f(\boldsymbol{\theta}-\frac{\pi}{2}\hat{e}_{A_i})$
and thus all of them cancel out.
One can simplify Eq. \eqref{delta theta_A expanded full} further into

\begin{equation}
\begin{split} \label{Delta theta_A before final approx}
        & \Delta \theta_{A_i} = \frac{\pi}{2} \Big[ \bra{\Phi_P}[\hat{B}, \hat{\mathcal{G}}_{A_i}]\ket{\Phi_P} \\
        & + \theta_{A_i} \bra{\Phi_P}\Big[ [\hat{B}, \hat{\mathcal{G}}_{A_i}], \hat{\mathcal{G}}_{A_i} \Big] \ket{\Phi_P} \\
        & + \frac{1}{2} \sum_{A_k \ne A_i} \theta_{A_k} \bra{\Phi_P}\Big[ [\hat{B}, \hat{\mathcal{G}}_{A_i}], \hat{\mathcal{G}}_{A_k} \Big] \ket{\Phi_P} +... \Big] 
\end{split}
\end{equation}

In Eq. \eqref{Delta theta_A before final approx} the terms involving \enquote{non-diagonal} double commutators with $\hat{\mathcal{G}}_{A_i}$
and $\hat{\mathcal{G}}_{A_k}$ are ignored since in general such terms are extremely small with respect to other leading order terms. Moreover, retaining such terms would lead to additional measurement overhead.
Combined with the adiabatic approximation, this provides us with
\begin{equation}
\begin{split}
        & \Delta \theta_{A_i} \approx \frac{\pi}{2} \Big[ \bra{\Phi_P}[\hat{B}, \hat{\mathcal{G}}_{A_i}]\ket{\Phi_P} \\
        & + \theta_{A_i} \bra{\Phi_P}\Big[ [\hat{B}, \hat{\mathcal{G}}_{A_i}], \hat{\mathcal{G}}_{A_i} \Big] \ket{\Phi_P} \Big] = 0
\end{split}
\end{equation}
which leads to the principal-to-auxiliary mapping functional (Eq. \eqref{theta_A as function of theta_P}).
Similar approximation is used to obtain the expression for the approximate cost function with ASC (Eq. \eqref{cost function with ASC}).

\section*{References:}
% \bibliography{./literature}

\begin{thebibliography}{57}%
\makeatletter
\providecommand \@ifxundefined [1]{%
 \@ifx{#1\undefined}
}%
\providecommand \@ifnum [1]{%
 \ifnum #1\expandafter \@firstoftwo
 \else \expandafter \@secondoftwo
 \fi
}%
\providecommand \@ifx [1]{%
 \ifx #1\expandafter \@firstoftwo
 \else \expandafter \@secondoftwo
 \fi
}%
\providecommand \natexlab [1]{#1}%
\providecommand \enquote  [1]{``#1''}%
\providecommand \bibnamefont  [1]{#1}%
\providecommand \bibfnamefont [1]{#1}%
\providecommand \citenamefont [1]{#1}%
\providecommand \href@noop [0]{\@secondoftwo}%
\providecommand \href [0]{\begingroup \@sanitize@url \@href}%
\providecommand \@href[1]{\@@startlink{#1}\@@href}%
\providecommand \@@href[1]{\endgroup#1\@@endlink}%
\providecommand \@sanitize@url [0]{\catcode `\\12\catcode `\$12\catcode `\&12\catcode `\#12\catcode `\^12\catcode `\_12\catcode `\%12\relax}%
\providecommand \@@startlink[1]{}%
\providecommand \@@endlink[0]{}%
\providecommand \url  [0]{\begingroup\@sanitize@url \@url }%
\providecommand \@url [1]{\endgroup\@href {#1}{\urlprefix }}%
\providecommand \urlprefix  [0]{URL }%
\providecommand \Eprint [0]{\href }%
\providecommand \doibase [0]{https://doi.org/}%
\providecommand \selectlanguage [0]{\@gobble}%
\providecommand \bibinfo  [0]{\@secondoftwo}%
\providecommand \bibfield  [0]{\@secondoftwo}%
\providecommand \translation [1]{[#1]}%
\providecommand \BibitemOpen [0]{}%
\providecommand \bibitemStop [0]{}%
\providecommand \bibitemNoStop [0]{.\EOS\space}%
\providecommand \EOS [0]{\spacefactor3000\relax}%
\providecommand \BibitemShut  [1]{\csname bibitem#1\endcsname}%
\let\auto@bib@innerbib\@empty
%</preamble>
\bibitem [{\citenamefont {Peruzzo}\ \emph {et~al.}(2014)\citenamefont {Peruzzo}, \citenamefont {McClean}, \citenamefont {Shadbolt}, \citenamefont {Yung}, \citenamefont {Zhou}, \citenamefont {Love}, \citenamefont {Aspuru-Guzik},\ and\ \citenamefont {O'brien}}]{peruzzo2014variational}%
  \BibitemOpen
  \bibfield  {author} {\bibinfo {author} {\bibfnamefont {A.}~\bibnamefont {Peruzzo}}, \bibinfo {author} {\bibfnamefont {J.}~\bibnamefont {McClean}}, \bibinfo {author} {\bibfnamefont {P.}~\bibnamefont {Shadbolt}}, \bibinfo {author} {\bibfnamefont {M.-H.}\ \bibnamefont {Yung}}, \bibinfo {author} {\bibfnamefont {X.-Q.}\ \bibnamefont {Zhou}}, \bibinfo {author} {\bibfnamefont {P.~J.}\ \bibnamefont {Love}}, \bibinfo {author} {\bibfnamefont {A.}~\bibnamefont {Aspuru-Guzik}},\ and\ \bibinfo {author} {\bibfnamefont {J.~L.}\ \bibnamefont {O'brien}},\ }\bibfield  {title} {\enquote {\bibinfo {title} {A variational eigenvalue solver on a photonic quantum processor},}\ }\href@noop {} {\bibfield  {journal} {\bibinfo  {journal} {Nature communications}\ }\textbf {\bibinfo {volume} {5}},\ \bibinfo {pages} {4213} (\bibinfo {year} {2014})}\BibitemShut {NoStop}%
\bibitem [{\citenamefont {Cerezo}\ \emph {et~al.}(2021{\natexlab{a}})\citenamefont {Cerezo}, \citenamefont {Arrasmith}, \citenamefont {Babbush}, \citenamefont {Benjamin}, \citenamefont {Endo}, \citenamefont {Fujii}, \citenamefont {McClean}, \citenamefont {Mitarai}, \citenamefont {Yuan}, \citenamefont {Cincio} \emph {et~al.}}]{cerezo2021variational}%
  \BibitemOpen
  \bibfield  {author} {\bibinfo {author} {\bibfnamefont {M.}~\bibnamefont {Cerezo}}, \bibinfo {author} {\bibfnamefont {A.}~\bibnamefont {Arrasmith}}, \bibinfo {author} {\bibfnamefont {R.}~\bibnamefont {Babbush}}, \bibinfo {author} {\bibfnamefont {S.~C.}\ \bibnamefont {Benjamin}}, \bibinfo {author} {\bibfnamefont {S.}~\bibnamefont {Endo}}, \bibinfo {author} {\bibfnamefont {K.}~\bibnamefont {Fujii}}, \bibinfo {author} {\bibfnamefont {J.~R.}\ \bibnamefont {McClean}}, \bibinfo {author} {\bibfnamefont {K.}~\bibnamefont {Mitarai}}, \bibinfo {author} {\bibfnamefont {X.}~\bibnamefont {Yuan}}, \bibinfo {author} {\bibfnamefont {L.}~\bibnamefont {Cincio}}, \emph {et~al.},\ }\bibfield  {title} {\enquote {\bibinfo {title} {Variational quantum algorithms},}\ }\href@noop {} {\bibfield  {journal} {\bibinfo  {journal} {Nature Reviews Physics}\ }\textbf {\bibinfo {volume} {3}},\ \bibinfo {pages} {625--644} (\bibinfo {year} {2021}{\natexlab{a}})}\BibitemShut {NoStop}%
\bibitem [{\citenamefont {Bharti}\ \emph {et~al.}(2022)\citenamefont {Bharti}, \citenamefont {Cervera-Lierta}, \citenamefont {Kyaw}, \citenamefont {Haug}, \citenamefont {Alperin-Lea}, \citenamefont {Anand}, \citenamefont {Degroote}, \citenamefont {Heimonen}, \citenamefont {Kottmann}, \citenamefont {Menke} \emph {et~al.}}]{bharti2022noisy}%
  \BibitemOpen
  \bibfield  {author} {\bibinfo {author} {\bibfnamefont {K.}~\bibnamefont {Bharti}}, \bibinfo {author} {\bibfnamefont {A.}~\bibnamefont {Cervera-Lierta}}, \bibinfo {author} {\bibfnamefont {T.~H.}\ \bibnamefont {Kyaw}}, \bibinfo {author} {\bibfnamefont {T.}~\bibnamefont {Haug}}, \bibinfo {author} {\bibfnamefont {S.}~\bibnamefont {Alperin-Lea}}, \bibinfo {author} {\bibfnamefont {A.}~\bibnamefont {Anand}}, \bibinfo {author} {\bibfnamefont {M.}~\bibnamefont {Degroote}}, \bibinfo {author} {\bibfnamefont {H.}~\bibnamefont {Heimonen}}, \bibinfo {author} {\bibfnamefont {J.~S.}\ \bibnamefont {Kottmann}}, \bibinfo {author} {\bibfnamefont {T.}~\bibnamefont {Menke}}, \emph {et~al.},\ }\bibfield  {title} {\enquote {\bibinfo {title} {Noisy intermediate-scale quantum algorithms},}\ }\href@noop {} {\bibfield  {journal} {\bibinfo  {journal} {Reviews of Modern Physics}\ }\textbf {\bibinfo {volume} {94}},\ \bibinfo {pages} {015004} (\bibinfo {year} {2022})}\BibitemShut {NoStop}%
\bibitem [{\citenamefont {McClean}\ \emph {et~al.}(2018)\citenamefont {McClean}, \citenamefont {Boixo}, \citenamefont {Smelyanskiy}, \citenamefont {Babbush},\ and\ \citenamefont {Neven}}]{mcclean2018barren}%
  \BibitemOpen
  \bibfield  {author} {\bibinfo {author} {\bibfnamefont {J.~R.}\ \bibnamefont {McClean}}, \bibinfo {author} {\bibfnamefont {S.}~\bibnamefont {Boixo}}, \bibinfo {author} {\bibfnamefont {V.~N.}\ \bibnamefont {Smelyanskiy}}, \bibinfo {author} {\bibfnamefont {R.}~\bibnamefont {Babbush}},\ and\ \bibinfo {author} {\bibfnamefont {H.}~\bibnamefont {Neven}},\ }\bibfield  {title} {\enquote {\bibinfo {title} {Barren plateaus in quantum neural network training landscapes},}\ }\href@noop {} {\bibfield  {journal} {\bibinfo  {journal} {Nature communications}\ }\textbf {\bibinfo {volume} {9}},\ \bibinfo {pages} {4812} (\bibinfo {year} {2018})}\BibitemShut {NoStop}%
\bibitem [{\citenamefont {Cerezo}\ \emph {et~al.}(2021{\natexlab{b}})\citenamefont {Cerezo}, \citenamefont {Sone}, \citenamefont {Volkoff}, \citenamefont {Cincio},\ and\ \citenamefont {Coles}}]{cerezo2021cost}%
  \BibitemOpen
  \bibfield  {author} {\bibinfo {author} {\bibfnamefont {M.}~\bibnamefont {Cerezo}}, \bibinfo {author} {\bibfnamefont {A.}~\bibnamefont {Sone}}, \bibinfo {author} {\bibfnamefont {T.}~\bibnamefont {Volkoff}}, \bibinfo {author} {\bibfnamefont {L.}~\bibnamefont {Cincio}},\ and\ \bibinfo {author} {\bibfnamefont {P.~J.}\ \bibnamefont {Coles}},\ }\bibfield  {title} {\enquote {\bibinfo {title} {Cost function dependent barren plateaus in shallow parametrized quantum circuits},}\ }\href@noop {} {\bibfield  {journal} {\bibinfo  {journal} {Nature communications}\ }\textbf {\bibinfo {volume} {12}},\ \bibinfo {pages} {1791} (\bibinfo {year} {2021}{\natexlab{b}})}\BibitemShut {NoStop}%
\bibitem [{\citenamefont {Larocca}\ \emph {et~al.}(2025)\citenamefont {Larocca}, \citenamefont {Thanasilp}, \citenamefont {Wang}, \citenamefont {Sharma}, \citenamefont {Biamonte}, \citenamefont {Coles}, \citenamefont {Cincio}, \citenamefont {McClean}, \citenamefont {Holmes},\ and\ \citenamefont {Cerezo}}]{larocca2025barren}%
  \BibitemOpen
  \bibfield  {author} {\bibinfo {author} {\bibfnamefont {M.}~\bibnamefont {Larocca}}, \bibinfo {author} {\bibfnamefont {S.}~\bibnamefont {Thanasilp}}, \bibinfo {author} {\bibfnamefont {S.}~\bibnamefont {Wang}}, \bibinfo {author} {\bibfnamefont {K.}~\bibnamefont {Sharma}}, \bibinfo {author} {\bibfnamefont {J.}~\bibnamefont {Biamonte}}, \bibinfo {author} {\bibfnamefont {P.~J.}\ \bibnamefont {Coles}}, \bibinfo {author} {\bibfnamefont {L.}~\bibnamefont {Cincio}}, \bibinfo {author} {\bibfnamefont {J.~R.}\ \bibnamefont {McClean}}, \bibinfo {author} {\bibfnamefont {Z.}~\bibnamefont {Holmes}},\ and\ \bibinfo {author} {\bibfnamefont {M.}~\bibnamefont {Cerezo}},\ }\bibfield  {title} {\enquote {\bibinfo {title} {Barren plateaus in variational quantum computing},}\ }\href@noop {} {\bibfield  {journal} {\bibinfo  {journal} {Nature Reviews Physics}\ ,\ \bibinfo {pages} {1--16}} (\bibinfo {year} {2025})}\BibitemShut {NoStop}%
\bibitem [{\citenamefont {Anschuetz}\ and\ \citenamefont {Kiani}(2022)}]{anschuetz2022quantum}%
  \BibitemOpen
  \bibfield  {author} {\bibinfo {author} {\bibfnamefont {E.~R.}\ \bibnamefont {Anschuetz}}\ and\ \bibinfo {author} {\bibfnamefont {B.~T.}\ \bibnamefont {Kiani}},\ }\bibfield  {title} {\enquote {\bibinfo {title} {Quantum variational algorithms are swamped with traps},}\ }\href@noop {} {\bibfield  {journal} {\bibinfo  {journal} {Nature Communications}\ }\textbf {\bibinfo {volume} {13}},\ \bibinfo {pages} {7760} (\bibinfo {year} {2022})}\BibitemShut {NoStop}%
\bibitem [{\citenamefont {Larocca}\ \emph {et~al.}(2023)\citenamefont {Larocca}, \citenamefont {Ju}, \citenamefont {Garc{\'\i}a-Mart{\'\i}n}, \citenamefont {Coles},\ and\ \citenamefont {Cerezo}}]{larocca2023theory}%
  \BibitemOpen
  \bibfield  {author} {\bibinfo {author} {\bibfnamefont {M.}~\bibnamefont {Larocca}}, \bibinfo {author} {\bibfnamefont {N.}~\bibnamefont {Ju}}, \bibinfo {author} {\bibfnamefont {D.}~\bibnamefont {Garc{\'\i}a-Mart{\'\i}n}}, \bibinfo {author} {\bibfnamefont {P.~J.}\ \bibnamefont {Coles}},\ and\ \bibinfo {author} {\bibfnamefont {M.}~\bibnamefont {Cerezo}},\ }\bibfield  {title} {\enquote {\bibinfo {title} {Theory of overparametrization in quantum neural networks},}\ }\href@noop {} {\bibfield  {journal} {\bibinfo  {journal} {Nature Computational Science}\ }\textbf {\bibinfo {volume} {3}},\ \bibinfo {pages} {542--551} (\bibinfo {year} {2023})}\BibitemShut {NoStop}%
\bibitem [{\citenamefont {Kiani}, \citenamefont {Lloyd},\ and\ \citenamefont {Maity}(2020)}]{kiani2020learning}%
  \BibitemOpen
  \bibfield  {author} {\bibinfo {author} {\bibfnamefont {B.~T.}\ \bibnamefont {Kiani}}, \bibinfo {author} {\bibfnamefont {S.}~\bibnamefont {Lloyd}},\ and\ \bibinfo {author} {\bibfnamefont {R.}~\bibnamefont {Maity}},\ }\bibfield  {title} {\enquote {\bibinfo {title} {Learning unitaries by gradient descent},}\ }\href@noop {} {\bibfield  {journal} {\bibinfo  {journal} {arXiv preprint arXiv:2001.11897}\ } (\bibinfo {year} {2020})}\BibitemShut {NoStop}%
\bibitem [{\citenamefont {Wiersema}\ \emph {et~al.}(2020)\citenamefont {Wiersema}, \citenamefont {Zhou}, \citenamefont {de~Sereville}, \citenamefont {Carrasquilla}, \citenamefont {Kim},\ and\ \citenamefont {Yuen}}]{wiersema2020exploring}%
  \BibitemOpen
  \bibfield  {author} {\bibinfo {author} {\bibfnamefont {R.}~\bibnamefont {Wiersema}}, \bibinfo {author} {\bibfnamefont {C.}~\bibnamefont {Zhou}}, \bibinfo {author} {\bibfnamefont {Y.}~\bibnamefont {de~Sereville}}, \bibinfo {author} {\bibfnamefont {J.~F.}\ \bibnamefont {Carrasquilla}}, \bibinfo {author} {\bibfnamefont {Y.~B.}\ \bibnamefont {Kim}},\ and\ \bibinfo {author} {\bibfnamefont {H.}~\bibnamefont {Yuen}},\ }\bibfield  {title} {\enquote {\bibinfo {title} {Exploring entanglement and optimization within the hamiltonian variational ansatz},}\ }\href@noop {} {\bibfield  {journal} {\bibinfo  {journal} {PRX quantum}\ }\textbf {\bibinfo {volume} {1}},\ \bibinfo {pages} {020319} (\bibinfo {year} {2020})}\BibitemShut {NoStop}%
\bibitem [{\citenamefont {Grimsley}\ \emph {et~al.}(2019)\citenamefont {Grimsley}, \citenamefont {Economou}, \citenamefont {Barnes},\ and\ \citenamefont {Mayhall}}]{grimsley2019adaptive}%
  \BibitemOpen
  \bibfield  {author} {\bibinfo {author} {\bibfnamefont {H.~R.}\ \bibnamefont {Grimsley}}, \bibinfo {author} {\bibfnamefont {S.~E.}\ \bibnamefont {Economou}}, \bibinfo {author} {\bibfnamefont {E.}~\bibnamefont {Barnes}},\ and\ \bibinfo {author} {\bibfnamefont {N.~J.}\ \bibnamefont {Mayhall}},\ }\bibfield  {title} {\enquote {\bibinfo {title} {An adaptive variational algorithm for exact molecular simulations on a quantum computer},}\ }\href@noop {} {\bibfield  {journal} {\bibinfo  {journal} {Nature communications}\ }\textbf {\bibinfo {volume} {10}},\ \bibinfo {pages} {3007} (\bibinfo {year} {2019})}\BibitemShut {NoStop}%
\bibitem [{\citenamefont {Grimsley}\ \emph {et~al.}(2023)\citenamefont {Grimsley}, \citenamefont {Barron}, \citenamefont {Barnes}, \citenamefont {Economou},\ and\ \citenamefont {Mayhall}}]{Grimsley2023}%
  \BibitemOpen
  \bibfield  {author} {\bibinfo {author} {\bibfnamefont {H.~R.}\ \bibnamefont {Grimsley}}, \bibinfo {author} {\bibfnamefont {G.~S.}\ \bibnamefont {Barron}}, \bibinfo {author} {\bibfnamefont {E.}~\bibnamefont {Barnes}}, \bibinfo {author} {\bibfnamefont {S.~E.}\ \bibnamefont {Economou}},\ and\ \bibinfo {author} {\bibfnamefont {N.~J.}\ \bibnamefont {Mayhall}},\ }\bibfield  {title} {\enquote {\bibinfo {title} {Adaptive, problem-tailored variational quantum eigensolver mitigates rough parameter landscapes and barren plateaus},}\ }\href@noop {} {\bibfield  {journal} {\bibinfo  {journal} {npj Quantum Information}\ }\textbf {\bibinfo {volume} {9}} (\bibinfo {year} {2023})}\BibitemShut {NoStop}%
\bibitem [{\citenamefont {Tang}\ \emph {et~al.}(2021)\citenamefont {Tang}, \citenamefont {Shkolnikov}, \citenamefont {Barron}, \citenamefont {Grimsley}, \citenamefont {Mayhall}, \citenamefont {Barnes},\ and\ \citenamefont {Economou}}]{tang2021qubit}%
  \BibitemOpen
  \bibfield  {author} {\bibinfo {author} {\bibfnamefont {H.~L.}\ \bibnamefont {Tang}}, \bibinfo {author} {\bibfnamefont {V.}~\bibnamefont {Shkolnikov}}, \bibinfo {author} {\bibfnamefont {G.~S.}\ \bibnamefont {Barron}}, \bibinfo {author} {\bibfnamefont {H.~R.}\ \bibnamefont {Grimsley}}, \bibinfo {author} {\bibfnamefont {N.~J.}\ \bibnamefont {Mayhall}}, \bibinfo {author} {\bibfnamefont {E.}~\bibnamefont {Barnes}},\ and\ \bibinfo {author} {\bibfnamefont {S.~E.}\ \bibnamefont {Economou}},\ }\bibfield  {title} {\enquote {\bibinfo {title} {qubit-adapt-vqe: An adaptive algorithm for constructing hardware-efficient ans{\"a}tze on a quantum processor},}\ }\href@noop {} {\bibfield  {journal} {\bibinfo  {journal} {PRX Quantum}\ }\textbf {\bibinfo {volume} {2}},\ \bibinfo {pages} {020310} (\bibinfo {year} {2021})}\BibitemShut {NoStop}%
\bibitem [{\citenamefont {Yordanov}\ \emph {et~al.}(2021)\citenamefont {Yordanov}, \citenamefont {Armaos}, \citenamefont {Barnes},\ and\ \citenamefont {Arvidsson-Shukur}}]{yordanov2021qubit}%
  \BibitemOpen
  \bibfield  {author} {\bibinfo {author} {\bibfnamefont {Y.~S.}\ \bibnamefont {Yordanov}}, \bibinfo {author} {\bibfnamefont {V.}~\bibnamefont {Armaos}}, \bibinfo {author} {\bibfnamefont {C.~H.}\ \bibnamefont {Barnes}},\ and\ \bibinfo {author} {\bibfnamefont {D.~R.}\ \bibnamefont {Arvidsson-Shukur}},\ }\bibfield  {title} {\enquote {\bibinfo {title} {Qubit-excitation-based adaptive variational quantum eigensolver},}\ }\href@noop {} {\bibfield  {journal} {\bibinfo  {journal} {Communications Physics}\ }\textbf {\bibinfo {volume} {4}},\ \bibinfo {pages} {228} (\bibinfo {year} {2021})}\BibitemShut {NoStop}%
\bibitem [{\citenamefont {Zhu}\ \emph {et~al.}(2022)\citenamefont {Zhu}, \citenamefont {Tang}, \citenamefont {Barron}, \citenamefont {Calderon-Vargas}, \citenamefont {Mayhall}, \citenamefont {Barnes},\ and\ \citenamefont {Economou}}]{zhu2022adaptive}%
  \BibitemOpen
  \bibfield  {author} {\bibinfo {author} {\bibfnamefont {L.}~\bibnamefont {Zhu}}, \bibinfo {author} {\bibfnamefont {H.~L.}\ \bibnamefont {Tang}}, \bibinfo {author} {\bibfnamefont {G.~S.}\ \bibnamefont {Barron}}, \bibinfo {author} {\bibfnamefont {F.}~\bibnamefont {Calderon-Vargas}}, \bibinfo {author} {\bibfnamefont {N.~J.}\ \bibnamefont {Mayhall}}, \bibinfo {author} {\bibfnamefont {E.}~\bibnamefont {Barnes}},\ and\ \bibinfo {author} {\bibfnamefont {S.~E.}\ \bibnamefont {Economou}},\ }\bibfield  {title} {\enquote {\bibinfo {title} {Adaptive quantum approximate optimization algorithm for solving combinatorial problems on a quantum computer},}\ }\href@noop {} {\bibfield  {journal} {\bibinfo  {journal} {Physical Review Research}\ }\textbf {\bibinfo {volume} {4}},\ \bibinfo {pages} {033029} (\bibinfo {year} {2022})}\BibitemShut {NoStop}%
\bibitem [{\citenamefont {Mondal}\ \emph {et~al.}(2023)\citenamefont {Mondal}, \citenamefont {Halder}, \citenamefont {Halder},\ and\ \citenamefont {Maitra}}]{mondal2023development}%
  \BibitemOpen
  \bibfield  {author} {\bibinfo {author} {\bibfnamefont {D.}~\bibnamefont {Mondal}}, \bibinfo {author} {\bibfnamefont {D.}~\bibnamefont {Halder}}, \bibinfo {author} {\bibfnamefont {S.}~\bibnamefont {Halder}},\ and\ \bibinfo {author} {\bibfnamefont {R.}~\bibnamefont {Maitra}},\ }\bibfield  {title} {\enquote {\bibinfo {title} {{Development of a compact Ansatz via operator commutativity screening: Digital quantum simulation of molecular systems}},}\ }\href {https://doi.org/10.1063/5.0153182} {\bibfield  {journal} {\bibinfo  {journal} {The Journal of Chemical Physics}\ }\textbf {\bibinfo {volume} {159}},\ \bibinfo {pages} {014105} (\bibinfo {year} {2023})}\BibitemShut {NoStop}%
\bibitem [{\citenamefont {Halder}, \citenamefont {Prasannaa},\ and\ \citenamefont {Maitra}(2022)}]{halder2022dual}%
  \BibitemOpen
  \bibfield  {author} {\bibinfo {author} {\bibfnamefont {D.}~\bibnamefont {Halder}}, \bibinfo {author} {\bibfnamefont {V.~S.}\ \bibnamefont {Prasannaa}},\ and\ \bibinfo {author} {\bibfnamefont {R.}~\bibnamefont {Maitra}},\ }\bibfield  {title} {\enquote {\bibinfo {title} {{Dual exponential coupled cluster theory: Unitary adaptation, implementation in the variational quantum eigensolver framework and pilot applications}},}\ }\href {https://doi.org/10.1063/5.0114688} {\bibfield  {journal} {\bibinfo  {journal} {The Journal of Chemical Physics}\ }\textbf {\bibinfo {volume} {157}},\ \bibinfo {pages} {174117} (\bibinfo {year} {2022})}\BibitemShut {NoStop}%
\bibitem [{\citenamefont {Stair}\ and\ \citenamefont {Evangelista}(2021)}]{stair2021simulating}%
  \BibitemOpen
  \bibfield  {author} {\bibinfo {author} {\bibfnamefont {N.~H.}\ \bibnamefont {Stair}}\ and\ \bibinfo {author} {\bibfnamefont {F.~A.}\ \bibnamefont {Evangelista}},\ }\bibfield  {title} {\enquote {\bibinfo {title} {Simulating many-body systems with a projective quantum eigensolver},}\ }\href@noop {} {\bibfield  {journal} {\bibinfo  {journal} {PRX Quantum}\ }\textbf {\bibinfo {volume} {2}},\ \bibinfo {pages} {030301} (\bibinfo {year} {2021})}\BibitemShut {NoStop}%
\bibitem [{\citenamefont {Ryabinkin}\ \emph {et~al.}(2018)\citenamefont {Ryabinkin}, \citenamefont {Yen}, \citenamefont {Genin},\ and\ \citenamefont {Izmaylov}}]{ryabinkin2018qubit}%
  \BibitemOpen
  \bibfield  {author} {\bibinfo {author} {\bibfnamefont {I.~G.}\ \bibnamefont {Ryabinkin}}, \bibinfo {author} {\bibfnamefont {T.-C.}\ \bibnamefont {Yen}}, \bibinfo {author} {\bibfnamefont {S.~N.}\ \bibnamefont {Genin}},\ and\ \bibinfo {author} {\bibfnamefont {A.~F.}\ \bibnamefont {Izmaylov}},\ }\bibfield  {title} {\enquote {\bibinfo {title} {Qubit coupled cluster method: a systematic approach to quantum chemistry on a quantum computer},}\ }\href@noop {} {\bibfield  {journal} {\bibinfo  {journal} {Journal of chemical theory and computation}\ }\textbf {\bibinfo {volume} {14}},\ \bibinfo {pages} {6317--6326} (\bibinfo {year} {2018})}\BibitemShut {NoStop}%
\bibitem [{\citenamefont {Ryabinkin}\ \emph {et~al.}(2020)\citenamefont {Ryabinkin}, \citenamefont {Lang}, \citenamefont {Genin},\ and\ \citenamefont {Izmaylov}}]{ryabinkin2020iterative}%
  \BibitemOpen
  \bibfield  {author} {\bibinfo {author} {\bibfnamefont {I.~G.}\ \bibnamefont {Ryabinkin}}, \bibinfo {author} {\bibfnamefont {R.~A.}\ \bibnamefont {Lang}}, \bibinfo {author} {\bibfnamefont {S.~N.}\ \bibnamefont {Genin}},\ and\ \bibinfo {author} {\bibfnamefont {A.~F.}\ \bibnamefont {Izmaylov}},\ }\bibfield  {title} {\enquote {\bibinfo {title} {Iterative qubit coupled cluster approach with efficient screening of generators},}\ }\href@noop {} {\bibfield  {journal} {\bibinfo  {journal} {Journal of chemical theory and computation}\ }\textbf {\bibinfo {volume} {16}},\ \bibinfo {pages} {1055--1063} (\bibinfo {year} {2020})}\BibitemShut {NoStop}%
\bibitem [{\citenamefont {Yordanov}, \citenamefont {Arvidsson-Shukur},\ and\ \citenamefont {Barnes}(2020)}]{yordanov2020efficient}%
  \BibitemOpen
  \bibfield  {author} {\bibinfo {author} {\bibfnamefont {Y.~S.}\ \bibnamefont {Yordanov}}, \bibinfo {author} {\bibfnamefont {D.~R.}\ \bibnamefont {Arvidsson-Shukur}},\ and\ \bibinfo {author} {\bibfnamefont {C.~H.}\ \bibnamefont {Barnes}},\ }\bibfield  {title} {\enquote {\bibinfo {title} {Efficient quantum circuits for quantum computational chemistry},}\ }\href@noop {} {\bibfield  {journal} {\bibinfo  {journal} {Physical Review A}\ }\textbf {\bibinfo {volume} {102}},\ \bibinfo {pages} {062612} (\bibinfo {year} {2020})}\BibitemShut {NoStop}%
\bibitem [{\citenamefont {Magoulas}\ and\ \citenamefont {Evangelista}(2023{\natexlab{a}})}]{magoulas2023cnot}%
  \BibitemOpen
  \bibfield  {author} {\bibinfo {author} {\bibfnamefont {I.}~\bibnamefont {Magoulas}}\ and\ \bibinfo {author} {\bibfnamefont {F.~A.}\ \bibnamefont {Evangelista}},\ }\bibfield  {title} {\enquote {\bibinfo {title} {Cnot-efficient circuits for arbitrary rank many-body fermionic and qubit excitations},}\ }\href@noop {} {\bibfield  {journal} {\bibinfo  {journal} {Journal of Chemical Theory and Computation}\ }\textbf {\bibinfo {volume} {19}},\ \bibinfo {pages} {822--836} (\bibinfo {year} {2023}{\natexlab{a}})}\BibitemShut {NoStop}%
\bibitem [{\citenamefont {Mondal}\ \emph {et~al.}(2024)\citenamefont {Mondal}, \citenamefont {Patra}, \citenamefont {Halder},\ and\ \citenamefont {Maitra}}]{mondal2024projective}%
  \BibitemOpen
  \bibfield  {author} {\bibinfo {author} {\bibfnamefont {D.}~\bibnamefont {Mondal}}, \bibinfo {author} {\bibfnamefont {C.}~\bibnamefont {Patra}}, \bibinfo {author} {\bibfnamefont {D.}~\bibnamefont {Halder}},\ and\ \bibinfo {author} {\bibfnamefont {R.}~\bibnamefont {Maitra}},\ }\bibfield  {title} {\enquote {\bibinfo {title} {Projective quantum eigensolver with generalized operators},}\ }\href@noop {} {\bibfield  {journal} {\bibinfo  {journal} {arXiv preprint arXiv:2410.16111}\ } (\bibinfo {year} {2024})}\BibitemShut {NoStop}%
\bibitem [{\citenamefont {Patra}\ \emph {et~al.}(2024)\citenamefont {Patra}, \citenamefont {Mukherjee}, \citenamefont {Halder}, \citenamefont {Mondal},\ and\ \citenamefont {Maitra}}]{patra2024toward}%
  \BibitemOpen
  \bibfield  {author} {\bibinfo {author} {\bibfnamefont {C.}~\bibnamefont {Patra}}, \bibinfo {author} {\bibfnamefont {D.}~\bibnamefont {Mukherjee}}, \bibinfo {author} {\bibfnamefont {S.}~\bibnamefont {Halder}}, \bibinfo {author} {\bibfnamefont {D.}~\bibnamefont {Mondal}},\ and\ \bibinfo {author} {\bibfnamefont {R.}~\bibnamefont {Maitra}},\ }\bibfield  {title} {\enquote {\bibinfo {title} {Toward a resource-optimized dynamic quantum algorithm via non-iterative auxiliary subspace corrections},}\ }\href@noop {} {\bibfield  {journal} {\bibinfo  {journal} {The Journal of Chemical Physics}\ }\textbf {\bibinfo {volume} {161}} (\bibinfo {year} {2024})}\BibitemShut {NoStop}%
\bibitem [{\citenamefont {Halder}\ \emph {et~al.}(2023{\natexlab{a}})\citenamefont {Halder}, \citenamefont {Halder}, \citenamefont {Mondal}, \citenamefont {Patra}, \citenamefont {Chakraborty},\ and\ \citenamefont {Maitra}}]{halder2023corrections}%
  \BibitemOpen
  \bibfield  {author} {\bibinfo {author} {\bibfnamefont {D.}~\bibnamefont {Halder}}, \bibinfo {author} {\bibfnamefont {S.}~\bibnamefont {Halder}}, \bibinfo {author} {\bibfnamefont {D.}~\bibnamefont {Mondal}}, \bibinfo {author} {\bibfnamefont {C.}~\bibnamefont {Patra}}, \bibinfo {author} {\bibfnamefont {A.}~\bibnamefont {Chakraborty}},\ and\ \bibinfo {author} {\bibfnamefont {R.}~\bibnamefont {Maitra}},\ }\bibfield  {title} {\enquote {\bibinfo {title} {Corrections beyond coupled cluster singles and doubles through selected generalized rank-two operators: digital quantum simulation of strongly correlated systems},}\ }\href@noop {} {\bibfield  {journal} {\bibinfo  {journal} {Journal of Chemical Sciences}\ }\textbf {\bibinfo {volume} {135}},\ \bibinfo {pages} {41} (\bibinfo {year} {2023}{\natexlab{a}})}\BibitemShut {NoStop}%
\bibitem [{\citenamefont {Halder}, \citenamefont {Mondal},\ and\ \citenamefont {Maitra}(2024)}]{halder2024noise}%
  \BibitemOpen
  \bibfield  {author} {\bibinfo {author} {\bibfnamefont {D.}~\bibnamefont {Halder}}, \bibinfo {author} {\bibfnamefont {D.}~\bibnamefont {Mondal}},\ and\ \bibinfo {author} {\bibfnamefont {R.}~\bibnamefont {Maitra}},\ }\bibfield  {title} {\enquote {\bibinfo {title} {Noise-independent route toward the genesis of a compact ansatz for molecular energetics: A dynamic approach},}\ }\href@noop {} {\bibfield  {journal} {\bibinfo  {journal} {The Journal of Chemical Physics}\ }\textbf {\bibinfo {volume} {160}} (\bibinfo {year} {2024})}\BibitemShut {NoStop}%
\bibitem [{\citenamefont {Halder}\ \emph {et~al.}(2024)\citenamefont {Halder}, \citenamefont {Dey}, \citenamefont {Shrikhande},\ and\ \citenamefont {Maitra}}]{halder2024machine}%
  \BibitemOpen
  \bibfield  {author} {\bibinfo {author} {\bibfnamefont {S.}~\bibnamefont {Halder}}, \bibinfo {author} {\bibfnamefont {A.}~\bibnamefont {Dey}}, \bibinfo {author} {\bibfnamefont {C.}~\bibnamefont {Shrikhande}},\ and\ \bibinfo {author} {\bibfnamefont {R.}~\bibnamefont {Maitra}},\ }\bibfield  {title} {\enquote {\bibinfo {title} {Machine learning assisted construction of a shallow depth dynamic ansatz for noisy quantum hardware},}\ }\href@noop {} {\bibfield  {journal} {\bibinfo  {journal} {Chemical Science}\ }\textbf {\bibinfo {volume} {15}},\ \bibinfo {pages} {3279--3289} (\bibinfo {year} {2024})}\BibitemShut {NoStop}%
\bibitem [{\citenamefont {Kowalski}(2021)}]{kowalski2021dimensionality}%
  \BibitemOpen
  \bibfield  {author} {\bibinfo {author} {\bibfnamefont {K.}~\bibnamefont {Kowalski}},\ }\bibfield  {title} {\enquote {\bibinfo {title} {Dimensionality reduction of the many-body problem using coupled-cluster subsystem flow equations: Classical and quantum computing perspective},}\ }\href@noop {} {\bibfield  {journal} {\bibinfo  {journal} {Physical Review A}\ }\textbf {\bibinfo {volume} {104}},\ \bibinfo {pages} {032804} (\bibinfo {year} {2021})}\BibitemShut {NoStop}%
\bibitem [{\citenamefont {Kowalski}\ and\ \citenamefont {Bauman}(2023)}]{kowalski2023quantum}%
  \BibitemOpen
  \bibfield  {author} {\bibinfo {author} {\bibfnamefont {K.}~\bibnamefont {Kowalski}}\ and\ \bibinfo {author} {\bibfnamefont {N.~P.}\ \bibnamefont {Bauman}},\ }\bibfield  {title} {\enquote {\bibinfo {title} {Quantum flow algorithms for simulating many-body systems on quantum computers},}\ }\href@noop {} {\bibfield  {journal} {\bibinfo  {journal} {Physical Review Letters}\ }\textbf {\bibinfo {volume} {131}},\ \bibinfo {pages} {200601} (\bibinfo {year} {2023})}\BibitemShut {NoStop}%
\bibitem [{\citenamefont {Robledo-Moreno}\ \emph {et~al.}(2024)\citenamefont {Robledo-Moreno}, \citenamefont {Motta}, \citenamefont {Haas}, \citenamefont {Javadi-Abhari}, \citenamefont {Jurcevic}, \citenamefont {Kirby}, \citenamefont {Martiel}, \citenamefont {Sharma}, \citenamefont {Sharma}, \citenamefont {Shirakawa} \emph {et~al.}}]{robledo2024chemistry}%
  \BibitemOpen
  \bibfield  {author} {\bibinfo {author} {\bibfnamefont {J.}~\bibnamefont {Robledo-Moreno}}, \bibinfo {author} {\bibfnamefont {M.}~\bibnamefont {Motta}}, \bibinfo {author} {\bibfnamefont {H.}~\bibnamefont {Haas}}, \bibinfo {author} {\bibfnamefont {A.}~\bibnamefont {Javadi-Abhari}}, \bibinfo {author} {\bibfnamefont {P.}~\bibnamefont {Jurcevic}}, \bibinfo {author} {\bibfnamefont {W.}~\bibnamefont {Kirby}}, \bibinfo {author} {\bibfnamefont {S.}~\bibnamefont {Martiel}}, \bibinfo {author} {\bibfnamefont {K.}~\bibnamefont {Sharma}}, \bibinfo {author} {\bibfnamefont {S.}~\bibnamefont {Sharma}}, \bibinfo {author} {\bibfnamefont {T.}~\bibnamefont {Shirakawa}}, \emph {et~al.},\ }\bibfield  {title} {\enquote {\bibinfo {title} {Chemistry beyond exact solutions on a quantum-centric supercomputer},}\ }\href@noop {} {\bibfield  {journal} {\bibinfo  {journal} {arXiv preprint arXiv:2405.05068}\ } (\bibinfo {year} {2024})}\BibitemShut {NoStop}%
\bibitem [{\citenamefont {Bittel}\ and\ \citenamefont {Kliesch}(2021)}]{bittel2021training}%
  \BibitemOpen
  \bibfield  {author} {\bibinfo {author} {\bibfnamefont {L.}~\bibnamefont {Bittel}}\ and\ \bibinfo {author} {\bibfnamefont {M.}~\bibnamefont {Kliesch}},\ }\bibfield  {title} {\enquote {\bibinfo {title} {Training variational quantum algorithms is np-hard},}\ }\href@noop {} {\bibfield  {journal} {\bibinfo  {journal} {Physical review letters}\ }\textbf {\bibinfo {volume} {127}},\ \bibinfo {pages} {120502} (\bibinfo {year} {2021})}\BibitemShut {NoStop}%
\bibitem [{\citenamefont {Holmes}\ \emph {et~al.}(2022)\citenamefont {Holmes}, \citenamefont {Sharma}, \citenamefont {Cerezo},\ and\ \citenamefont {Coles}}]{holmes2022connecting}%
  \BibitemOpen
  \bibfield  {author} {\bibinfo {author} {\bibfnamefont {Z.}~\bibnamefont {Holmes}}, \bibinfo {author} {\bibfnamefont {K.}~\bibnamefont {Sharma}}, \bibinfo {author} {\bibfnamefont {M.}~\bibnamefont {Cerezo}},\ and\ \bibinfo {author} {\bibfnamefont {P.~J.}\ \bibnamefont {Coles}},\ }\bibfield  {title} {\enquote {\bibinfo {title} {Connecting ansatz expressibility to gradient magnitudes and barren plateaus},}\ }\href@noop {} {\bibfield  {journal} {\bibinfo  {journal} {PRX Quantum}\ }\textbf {\bibinfo {volume} {3}},\ \bibinfo {pages} {010313} (\bibinfo {year} {2022})}\BibitemShut {NoStop}%
\bibitem [{\citenamefont {Grant}\ \emph {et~al.}(2019)\citenamefont {Grant}, \citenamefont {Wossnig}, \citenamefont {Ostaszewski},\ and\ \citenamefont {Benedetti}}]{grant2019initialization}%
  \BibitemOpen
  \bibfield  {author} {\bibinfo {author} {\bibfnamefont {E.}~\bibnamefont {Grant}}, \bibinfo {author} {\bibfnamefont {L.}~\bibnamefont {Wossnig}}, \bibinfo {author} {\bibfnamefont {M.}~\bibnamefont {Ostaszewski}},\ and\ \bibinfo {author} {\bibfnamefont {M.}~\bibnamefont {Benedetti}},\ }\bibfield  {title} {\enquote {\bibinfo {title} {An initialization strategy for addressing barren plateaus in parametrized quantum circuits},}\ }\href@noop {} {\bibfield  {journal} {\bibinfo  {journal} {Quantum}\ }\textbf {\bibinfo {volume} {3}},\ \bibinfo {pages} {214} (\bibinfo {year} {2019})}\BibitemShut {NoStop}%
\bibitem [{\citenamefont {Wang}\ \emph {et~al.}(2024)\citenamefont {Wang}, \citenamefont {Qi}, \citenamefont {Ferrie},\ and\ \citenamefont {Dong}}]{wang2024trainability}%
  \BibitemOpen
  \bibfield  {author} {\bibinfo {author} {\bibfnamefont {Y.}~\bibnamefont {Wang}}, \bibinfo {author} {\bibfnamefont {B.}~\bibnamefont {Qi}}, \bibinfo {author} {\bibfnamefont {C.}~\bibnamefont {Ferrie}},\ and\ \bibinfo {author} {\bibfnamefont {D.}~\bibnamefont {Dong}},\ }\bibfield  {title} {\enquote {\bibinfo {title} {Trainability enhancement of parameterized quantum circuits via reduced-domain parameter initialization},}\ }\href@noop {} {\bibfield  {journal} {\bibinfo  {journal} {Physical Review Applied}\ }\textbf {\bibinfo {volume} {22}},\ \bibinfo {pages} {054005} (\bibinfo {year} {2024})}\BibitemShut {NoStop}%
\bibitem [{\citenamefont {Patra}, \citenamefont {Halder},\ and\ \citenamefont {Maitra}(2024)}]{patra2024projective}%
  \BibitemOpen
  \bibfield  {author} {\bibinfo {author} {\bibfnamefont {C.}~\bibnamefont {Patra}}, \bibinfo {author} {\bibfnamefont {S.}~\bibnamefont {Halder}},\ and\ \bibinfo {author} {\bibfnamefont {R.}~\bibnamefont {Maitra}},\ }\bibfield  {title} {\enquote {\bibinfo {title} {Projective quantum eigensolver via adiabatically decoupled subsystem evolution: A resource efficient approach to molecular energetics in noisy quantum computers},}\ }\href@noop {} {\bibfield  {journal} {\bibinfo  {journal} {The Journal of Chemical Physics}\ }\textbf {\bibinfo {volume} {160}} (\bibinfo {year} {2024})}\BibitemShut {NoStop}%
\bibitem [{\citenamefont {Halder}\ \emph {et~al.}(2023{\natexlab{b}})\citenamefont {Halder}, \citenamefont {Patra}, \citenamefont {Mondal},\ and\ \citenamefont {Maitra}}]{halder2023machine}%
  \BibitemOpen
  \bibfield  {author} {\bibinfo {author} {\bibfnamefont {S.}~\bibnamefont {Halder}}, \bibinfo {author} {\bibfnamefont {C.}~\bibnamefont {Patra}}, \bibinfo {author} {\bibfnamefont {D.}~\bibnamefont {Mondal}},\ and\ \bibinfo {author} {\bibfnamefont {R.}~\bibnamefont {Maitra}},\ }\bibfield  {title} {\enquote {\bibinfo {title} {Machine learning aided dimensionality reduction toward a resource efficient projective quantum eigensolver: Formal development and pilot applications},}\ }\href@noop {} {\bibfield  {journal} {\bibinfo  {journal} {The Journal of Chemical Physics}\ }\textbf {\bibinfo {volume} {158}} (\bibinfo {year} {2023}{\natexlab{b}})}\BibitemShut {NoStop}%
\bibitem [{\citenamefont {Agarawal}, \citenamefont {Chakraborty},\ and\ \citenamefont {Maitra}(2020)}]{agarawal2020stability}%
  \BibitemOpen
  \bibfield  {author} {\bibinfo {author} {\bibfnamefont {V.}~\bibnamefont {Agarawal}}, \bibinfo {author} {\bibfnamefont {A.}~\bibnamefont {Chakraborty}},\ and\ \bibinfo {author} {\bibfnamefont {R.}~\bibnamefont {Maitra}},\ }\bibfield  {title} {\enquote {\bibinfo {title} {Stability analysis of a double similarity transformed coupled cluster theory},}\ }\href@noop {} {\bibfield  {journal} {\bibinfo  {journal} {The Journal of Chemical Physics}\ }\textbf {\bibinfo {volume} {153}},\ \bibinfo {pages} {084113} (\bibinfo {year} {2020})}\BibitemShut {NoStop}%
\bibitem [{\citenamefont {Agarawal}\ \emph {et~al.}(2021)\citenamefont {Agarawal}, \citenamefont {Roy}, \citenamefont {Chakraborty},\ and\ \citenamefont {Maitra}}]{agarawal2021accelerating}%
  \BibitemOpen
  \bibfield  {author} {\bibinfo {author} {\bibfnamefont {V.}~\bibnamefont {Agarawal}}, \bibinfo {author} {\bibfnamefont {S.}~\bibnamefont {Roy}}, \bibinfo {author} {\bibfnamefont {A.}~\bibnamefont {Chakraborty}},\ and\ \bibinfo {author} {\bibfnamefont {R.}~\bibnamefont {Maitra}},\ }\bibfield  {title} {\enquote {\bibinfo {title} {Accelerating coupled cluster calculations with nonlinear dynamics and supervised machine learning},}\ }\href@noop {} {\bibfield  {journal} {\bibinfo  {journal} {The Journal of Chemical Physics}\ }\textbf {\bibinfo {volume} {154}},\ \bibinfo {pages} {044110} (\bibinfo {year} {2021})}\BibitemShut {NoStop}%
\bibitem [{\citenamefont {Agarawal}, \citenamefont {Patra},\ and\ \citenamefont {Maitra}(2021)}]{agarawal2021approximate}%
  \BibitemOpen
  \bibfield  {author} {\bibinfo {author} {\bibfnamefont {V.}~\bibnamefont {Agarawal}}, \bibinfo {author} {\bibfnamefont {C.}~\bibnamefont {Patra}},\ and\ \bibinfo {author} {\bibfnamefont {R.}~\bibnamefont {Maitra}},\ }\bibfield  {title} {\enquote {\bibinfo {title} {An approximate coupled cluster theory via nonlinear dynamics and synergetics: The adiabatic decoupling conditions},}\ }\href@noop {} {\bibfield  {journal} {\bibinfo  {journal} {The Journal of Chemical Physics}\ }\textbf {\bibinfo {volume} {155}},\ \bibinfo {pages} {124115} (\bibinfo {year} {2021})}\BibitemShut {NoStop}%
\bibitem [{\citenamefont {Patra}\ \emph {et~al.}(2023)\citenamefont {Patra}, \citenamefont {Agarawal}, \citenamefont {Halder}, \citenamefont {Chakraborty}, \citenamefont {Mondal}, \citenamefont {Halder},\ and\ \citenamefont {Maitra}}]{patra2023synergistic}%
  \BibitemOpen
  \bibfield  {author} {\bibinfo {author} {\bibfnamefont {C.}~\bibnamefont {Patra}}, \bibinfo {author} {\bibfnamefont {V.}~\bibnamefont {Agarawal}}, \bibinfo {author} {\bibfnamefont {D.}~\bibnamefont {Halder}}, \bibinfo {author} {\bibfnamefont {A.}~\bibnamefont {Chakraborty}}, \bibinfo {author} {\bibfnamefont {D.}~\bibnamefont {Mondal}}, \bibinfo {author} {\bibfnamefont {S.}~\bibnamefont {Halder}},\ and\ \bibinfo {author} {\bibfnamefont {R.}~\bibnamefont {Maitra}},\ }\bibfield  {title} {\enquote {\bibinfo {title} {A synergistic approach towards optimization of coupled cluster amplitudes by exploiting dynamical hierarchy},}\ }\href@noop {} {\bibfield  {journal} {\bibinfo  {journal} {ChemPhysChem}\ }\textbf {\bibinfo {volume} {24}},\ \bibinfo {pages} {e202200633} (\bibinfo {year} {2023})}\BibitemShut {NoStop}%
\bibitem [{\citenamefont {Mitarai}\ \emph {et~al.}(2018)\citenamefont {Mitarai}, \citenamefont {Negoro}, \citenamefont {Kitagawa},\ and\ \citenamefont {Fujii}}]{mitarai2018quantum}%
  \BibitemOpen
  \bibfield  {author} {\bibinfo {author} {\bibfnamefont {K.}~\bibnamefont {Mitarai}}, \bibinfo {author} {\bibfnamefont {M.}~\bibnamefont {Negoro}}, \bibinfo {author} {\bibfnamefont {M.}~\bibnamefont {Kitagawa}},\ and\ \bibinfo {author} {\bibfnamefont {K.}~\bibnamefont {Fujii}},\ }\bibfield  {title} {\enquote {\bibinfo {title} {Quantum circuit learning},}\ }\href@noop {} {\bibfield  {journal} {\bibinfo  {journal} {Physical Review A}\ }\textbf {\bibinfo {volume} {98}},\ \bibinfo {pages} {032309} (\bibinfo {year} {2018})}\BibitemShut {NoStop}%
\bibitem [{\citenamefont {Schuld}\ \emph {et~al.}(2019)\citenamefont {Schuld}, \citenamefont {Bergholm}, \citenamefont {Gogolin}, \citenamefont {Izaac},\ and\ \citenamefont {Killoran}}]{schuld2019evaluating}%
  \BibitemOpen
  \bibfield  {author} {\bibinfo {author} {\bibfnamefont {M.}~\bibnamefont {Schuld}}, \bibinfo {author} {\bibfnamefont {V.}~\bibnamefont {Bergholm}}, \bibinfo {author} {\bibfnamefont {C.}~\bibnamefont {Gogolin}}, \bibinfo {author} {\bibfnamefont {J.}~\bibnamefont {Izaac}},\ and\ \bibinfo {author} {\bibfnamefont {N.}~\bibnamefont {Killoran}},\ }\bibfield  {title} {\enquote {\bibinfo {title} {Evaluating analytic gradients on quantum hardware},}\ }\href@noop {} {\bibfield  {journal} {\bibinfo  {journal} {Physical Review A}\ }\textbf {\bibinfo {volume} {99}},\ \bibinfo {pages} {032331} (\bibinfo {year} {2019})}\BibitemShut {NoStop}%
\bibitem [{\citenamefont {Wierichs}\ \emph {et~al.}(2022)\citenamefont {Wierichs}, \citenamefont {Izaac}, \citenamefont {Wang},\ and\ \citenamefont {Lin}}]{wierichs2022general}%
  \BibitemOpen
  \bibfield  {author} {\bibinfo {author} {\bibfnamefont {D.}~\bibnamefont {Wierichs}}, \bibinfo {author} {\bibfnamefont {J.}~\bibnamefont {Izaac}}, \bibinfo {author} {\bibfnamefont {C.}~\bibnamefont {Wang}},\ and\ \bibinfo {author} {\bibfnamefont {C.~Y.-Y.}\ \bibnamefont {Lin}},\ }\bibfield  {title} {\enquote {\bibinfo {title} {General parameter-shift rules for quantum gradients},}\ }\href@noop {} {\bibfield  {journal} {\bibinfo  {journal} {Quantum}\ }\textbf {\bibinfo {volume} {6}},\ \bibinfo {pages} {677} (\bibinfo {year} {2022})}\BibitemShut {NoStop}%
\bibitem [{\citenamefont {Van~Kampen}(1985)}]{van1985elimination}%
  \BibitemOpen
  \bibfield  {author} {\bibinfo {author} {\bibfnamefont {N.~G.}\ \bibnamefont {Van~Kampen}},\ }\bibfield  {title} {\enquote {\bibinfo {title} {Elimination of fast variables},}\ }\href@noop {} {\bibfield  {journal} {\bibinfo  {journal} {Physics Reports}\ }\textbf {\bibinfo {volume} {124}},\ \bibinfo {pages} {69--160} (\bibinfo {year} {1985})}\BibitemShut {NoStop}%
\bibitem [{\citenamefont {Claudino}\ \emph {et~al.}(2021)\citenamefont {Claudino}, \citenamefont {Peng}, \citenamefont {Bauman}, \citenamefont {Kowalski},\ and\ \citenamefont {Humble}}]{claudino2021improving}%
  \BibitemOpen
  \bibfield  {author} {\bibinfo {author} {\bibfnamefont {D.}~\bibnamefont {Claudino}}, \bibinfo {author} {\bibfnamefont {B.}~\bibnamefont {Peng}}, \bibinfo {author} {\bibfnamefont {N.~P.}\ \bibnamefont {Bauman}}, \bibinfo {author} {\bibfnamefont {K.}~\bibnamefont {Kowalski}},\ and\ \bibinfo {author} {\bibfnamefont {T.~S.}\ \bibnamefont {Humble}},\ }\bibfield  {title} {\enquote {\bibinfo {title} {Improving the accuracy and efficiency of quantum connected moments expansions},}\ }\href@noop {} {\bibfield  {journal} {\bibinfo  {journal} {Quantum Science and Technology}\ }\textbf {\bibinfo {volume} {6}},\ \bibinfo {pages} {034012} (\bibinfo {year} {2021})}\BibitemShut {NoStop}%
\bibitem [{\citenamefont {Magoulas}\ and\ \citenamefont {Evangelista}(2023{\natexlab{b}})}]{magoulas2023unitary}%
  \BibitemOpen
  \bibfield  {author} {\bibinfo {author} {\bibfnamefont {I.}~\bibnamefont {Magoulas}}\ and\ \bibinfo {author} {\bibfnamefont {F.~A.}\ \bibnamefont {Evangelista}},\ }\bibfield  {title} {\enquote {\bibinfo {title} {Unitary coupled cluster: Seizing the quantum moment},}\ }\href@noop {} {\bibfield  {journal} {\bibinfo  {journal} {The Journal of Physical Chemistry A}\ }\textbf {\bibinfo {volume} {127}},\ \bibinfo {pages} {6567--6576} (\bibinfo {year} {2023}{\natexlab{b}})}\BibitemShut {NoStop}%
\bibitem [{\citenamefont {Windom}, \citenamefont {Claudino},\ and\ \citenamefont {Bartlett}(2024)}]{windom2024new}%
  \BibitemOpen
  \bibfield  {author} {\bibinfo {author} {\bibfnamefont {Z.~W.}\ \bibnamefont {Windom}}, \bibinfo {author} {\bibfnamefont {D.}~\bibnamefont {Claudino}},\ and\ \bibinfo {author} {\bibfnamefont {R.~J.}\ \bibnamefont {Bartlett}},\ }\bibfield  {title} {\enquote {\bibinfo {title} {A new "gold standard": Perturbative triples corrections in unitary coupled cluster theory and prospects for quantum computing},}\ }\href@noop {} {\bibfield  {journal} {\bibinfo  {journal} {The Journal of Chemical Physics}\ }\textbf {\bibinfo {volume} {160}} (\bibinfo {year} {2024})}\BibitemShut {NoStop}%
\bibitem [{\citenamefont {Haidar}\ \emph {et~al.}(2024)\citenamefont {Haidar}, \citenamefont {Adjoua}, \citenamefont {Badreddine}, \citenamefont {Peruzzo},\ and\ \citenamefont {Piquemal}}]{haidar2024non}%
  \BibitemOpen
  \bibfield  {author} {\bibinfo {author} {\bibfnamefont {M.}~\bibnamefont {Haidar}}, \bibinfo {author} {\bibfnamefont {O.}~\bibnamefont {Adjoua}}, \bibinfo {author} {\bibfnamefont {S.}~\bibnamefont {Badreddine}}, \bibinfo {author} {\bibfnamefont {A.}~\bibnamefont {Peruzzo}},\ and\ \bibinfo {author} {\bibfnamefont {J.-P.}\ \bibnamefont {Piquemal}},\ }\bibfield  {title} {\enquote {\bibinfo {title} {Non-iterative disentangled unitary coupled-cluster based on lie-algebraic structure},}\ }\href@noop {} {\bibfield  {journal} {\bibinfo  {journal} {Quantum Science and Technology}\ } (\bibinfo {year} {2024})}\BibitemShut {NoStop}%
\bibitem [{\citenamefont {Kowalski}(2018)}]{kowalski2018properties}%
  \BibitemOpen
  \bibfield  {author} {\bibinfo {author} {\bibfnamefont {K.}~\bibnamefont {Kowalski}},\ }\bibfield  {title} {\enquote {\bibinfo {title} {Properties of coupled-cluster equations originating in excitation sub-algebras},}\ }\href@noop {} {\bibfield  {journal} {\bibinfo  {journal} {The Journal of Chemical Physics}\ }\textbf {\bibinfo {volume} {148}},\ \bibinfo {pages} {094104} (\bibinfo {year} {2018})}\BibitemShut {NoStop}%
\bibitem [{\citenamefont {Kowalski}, \citenamefont {Brabec},\ and\ \citenamefont {Peng}(2018)}]{kowalski2018regularized}%
  \BibitemOpen
  \bibfield  {author} {\bibinfo {author} {\bibfnamefont {K.}~\bibnamefont {Kowalski}}, \bibinfo {author} {\bibfnamefont {J.}~\bibnamefont {Brabec}},\ and\ \bibinfo {author} {\bibfnamefont {B.}~\bibnamefont {Peng}},\ }\bibfield  {title} {\enquote {\bibinfo {title} {Regularized and renormalized many-body techniques for describing correlated molecular systems: A {C}oupled-{C}luster perspective},}\ }\href@noop {} {\bibfield  {journal} {\bibinfo  {journal} {Annual Reports in Computational Chemistry}\ }\textbf {\bibinfo {volume} {14}},\ \bibinfo {pages} {3--45} (\bibinfo {year} {2018})}\BibitemShut {NoStop}%
\bibitem [{\citenamefont {Evangelista}, \citenamefont {Chan},\ and\ \citenamefont {Scuseria}(2019)}]{dUCC_evangelista}%
  \BibitemOpen
  \bibfield  {author} {\bibinfo {author} {\bibfnamefont {F.~A.}\ \bibnamefont {Evangelista}}, \bibinfo {author} {\bibfnamefont {G.~K.-L.}\ \bibnamefont {Chan}},\ and\ \bibinfo {author} {\bibfnamefont {G.~E.}\ \bibnamefont {Scuseria}},\ }\bibfield  {title} {\enquote {\bibinfo {title} {{Exact parameterization of fermionic wave functions via unitary coupled cluster theory}},}\ }\href {https://doi.org/10.1063/1.5133059} {\bibfield  {journal} {\bibinfo  {journal} {The Journal of Chemical Physics}\ }\textbf {\bibinfo {volume} {151}},\ \bibinfo {pages} {244112} (\bibinfo {year} {2019})},\ \Eprint {https://arxiv.org/abs/https://pubs.aip.org/aip/jcp/article-pdf/doi/10.1063/1.5133059/16659468/244112\_1\_online.pdf} {https://pubs.aip.org/aip/jcp/article-pdf/doi/10.1063/1.5133059/16659468/244112\_1\_online.pdf} \BibitemShut {NoStop}%
\bibitem [{\citenamefont {Javadi-Abhari}\ \emph {et~al.}(2024)\citenamefont {Javadi-Abhari}, \citenamefont {Treinish}, \citenamefont {Krsulich}, \citenamefont {Wood}, \citenamefont {Lishman}, \citenamefont {Gacon}, \citenamefont {Martiel}, \citenamefont {Nation}, \citenamefont {Bishop}, \citenamefont {Cross}, \citenamefont {Johnson},\ and\ \citenamefont {Gambetta}}]{qiskit2024}%
  \BibitemOpen
  \bibfield  {author} {\bibinfo {author} {\bibfnamefont {A.}~\bibnamefont {Javadi-Abhari}}, \bibinfo {author} {\bibfnamefont {M.}~\bibnamefont {Treinish}}, \bibinfo {author} {\bibfnamefont {K.}~\bibnamefont {Krsulich}}, \bibinfo {author} {\bibfnamefont {C.~J.}\ \bibnamefont {Wood}}, \bibinfo {author} {\bibfnamefont {J.}~\bibnamefont {Lishman}}, \bibinfo {author} {\bibfnamefont {J.}~\bibnamefont {Gacon}}, \bibinfo {author} {\bibfnamefont {S.}~\bibnamefont {Martiel}}, \bibinfo {author} {\bibfnamefont {P.~D.}\ \bibnamefont {Nation}}, \bibinfo {author} {\bibfnamefont {L.~S.}\ \bibnamefont {Bishop}}, \bibinfo {author} {\bibfnamefont {A.~W.}\ \bibnamefont {Cross}}, \bibinfo {author} {\bibfnamefont {B.~R.}\ \bibnamefont {Johnson}},\ and\ \bibinfo {author} {\bibfnamefont {J.~M.}\ \bibnamefont {Gambetta}},\ }\href {https://doi.org/10.48550/arXiv.2405.08810} {\enquote {\bibinfo {title} {Quantum computing with {Q}iskit},}\ } (\bibinfo {year} {2024}),\ \Eprint {https://arxiv.org/abs/2405.08810} {arXiv:2405.08810
  [quant-ph]} \BibitemShut {NoStop}%
\bibitem [{\citenamefont {Sun}\ \emph {et~al.}(2018)\citenamefont {Sun}, \citenamefont {Berkelbach}, \citenamefont {Blunt}, \citenamefont {Booth}, \citenamefont {Guo}, \citenamefont {Li}, \citenamefont {Liu}, \citenamefont {McClain}, \citenamefont {Sayfutyarova}, \citenamefont {Sharma} \emph {et~al.}}]{sun2018pyscf}%
  \BibitemOpen
  \bibfield  {author} {\bibinfo {author} {\bibfnamefont {Q.}~\bibnamefont {Sun}}, \bibinfo {author} {\bibfnamefont {T.~C.}\ \bibnamefont {Berkelbach}}, \bibinfo {author} {\bibfnamefont {N.~S.}\ \bibnamefont {Blunt}}, \bibinfo {author} {\bibfnamefont {G.~H.}\ \bibnamefont {Booth}}, \bibinfo {author} {\bibfnamefont {S.}~\bibnamefont {Guo}}, \bibinfo {author} {\bibfnamefont {Z.}~\bibnamefont {Li}}, \bibinfo {author} {\bibfnamefont {J.}~\bibnamefont {Liu}}, \bibinfo {author} {\bibfnamefont {J.~D.}\ \bibnamefont {McClain}}, \bibinfo {author} {\bibfnamefont {E.~R.}\ \bibnamefont {Sayfutyarova}}, \bibinfo {author} {\bibfnamefont {S.}~\bibnamefont {Sharma}}, \emph {et~al.},\ }\bibfield  {title} {\enquote {\bibinfo {title} {Pyscf: the python-based simulations of chemistry framework},}\ }\href@noop {} {\bibfield  {journal} {\bibinfo  {journal} {Wiley Interdisciplinary Reviews: Computational Molecular Science}\ }\textbf {\bibinfo {volume} {8}},\ \bibinfo {pages} {e1340} (\bibinfo {year} {2018})}\BibitemShut {NoStop}%
\bibitem [{\citenamefont {Giurgica-Tiron}\ \emph {et~al.}(2020)\citenamefont {Giurgica-Tiron}, \citenamefont {Hindy}, \citenamefont {LaRose}, \citenamefont {Mari},\ and\ \citenamefont {Zeng}}]{giurgica2020digital}%
  \BibitemOpen
  \bibfield  {author} {\bibinfo {author} {\bibfnamefont {T.}~\bibnamefont {Giurgica-Tiron}}, \bibinfo {author} {\bibfnamefont {Y.}~\bibnamefont {Hindy}}, \bibinfo {author} {\bibfnamefont {R.}~\bibnamefont {LaRose}}, \bibinfo {author} {\bibfnamefont {A.}~\bibnamefont {Mari}},\ and\ \bibinfo {author} {\bibfnamefont {W.~J.}\ \bibnamefont {Zeng}},\ }\bibfield  {title} {\enquote {\bibinfo {title} {Digital zero noise extrapolation for quantum error mitigation},}\ }in\ \href@noop {} {\emph {\bibinfo {booktitle} {2020 IEEE International Conference on Quantum Computing and Engineering (QCE)}}}\ (\bibinfo {organization} {IEEE},\ \bibinfo {year} {2020})\ pp.\ \bibinfo {pages} {306--316}\BibitemShut {NoStop}%
\bibitem [{\citenamefont {Zhou}\ \emph {et~al.}(2020)\citenamefont {Zhou}, \citenamefont {Wang}, \citenamefont {Choi}, \citenamefont {Pichler},\ and\ \citenamefont {Lukin}}]{zhou2020quantum}%
  \BibitemOpen
  \bibfield  {author} {\bibinfo {author} {\bibfnamefont {L.}~\bibnamefont {Zhou}}, \bibinfo {author} {\bibfnamefont {S.-T.}\ \bibnamefont {Wang}}, \bibinfo {author} {\bibfnamefont {S.}~\bibnamefont {Choi}}, \bibinfo {author} {\bibfnamefont {H.}~\bibnamefont {Pichler}},\ and\ \bibinfo {author} {\bibfnamefont {M.~D.}\ \bibnamefont {Lukin}},\ }\bibfield  {title} {\enquote {\bibinfo {title} {Quantum approximate optimization algorithm: Performance, mechanism, and implementation on near-term devices},}\ }\href@noop {} {\bibfield  {journal} {\bibinfo  {journal} {Physical Review X}\ }\textbf {\bibinfo {volume} {10}},\ \bibinfo {pages} {021067} (\bibinfo {year} {2020})}\BibitemShut {NoStop}%
\bibitem [{\citenamefont {Wierichs}, \citenamefont {Gogolin},\ and\ \citenamefont {Kastoryano}(2020)}]{wierichs2020avoiding}%
  \BibitemOpen
  \bibfield  {author} {\bibinfo {author} {\bibfnamefont {D.}~\bibnamefont {Wierichs}}, \bibinfo {author} {\bibfnamefont {C.}~\bibnamefont {Gogolin}},\ and\ \bibinfo {author} {\bibfnamefont {M.}~\bibnamefont {Kastoryano}},\ }\bibfield  {title} {\enquote {\bibinfo {title} {Avoiding local minima in variational quantum eigensolvers with the natural gradient optimizer},}\ }\href@noop {} {\bibfield  {journal} {\bibinfo  {journal} {Physical Review Research}\ }\textbf {\bibinfo {volume} {2}},\ \bibinfo {pages} {043246} (\bibinfo {year} {2020})}\BibitemShut {NoStop}%
\bibitem [{\citenamefont {Mao}, \citenamefont {Tian},\ and\ \citenamefont {Sun}(2024)}]{mao2024towards}%
  \BibitemOpen
  \bibfield  {author} {\bibinfo {author} {\bibfnamefont {R.}~\bibnamefont {Mao}}, \bibinfo {author} {\bibfnamefont {G.}~\bibnamefont {Tian}},\ and\ \bibinfo {author} {\bibfnamefont {X.}~\bibnamefont {Sun}},\ }\bibfield  {title} {\enquote {\bibinfo {title} {Towards determining the presence of barren plateaus in some chemically inspired variational quantum algorithms},}\ }\href@noop {} {\bibfield  {journal} {\bibinfo  {journal} {Communications Physics}\ }\textbf {\bibinfo {volume} {7}},\ \bibinfo {pages} {342} (\bibinfo {year} {2024})}\BibitemShut {NoStop}%
\end{thebibliography}

%aipnum4-2.bst 2019-01-14 (MD) hand-edited version of apsrev4-1.bst

\end{document}